\documentclass[twocolumn]{aastex631}
\pdfoutput=1

\usepackage[separate-uncertainty=true, range-phrase=-, range-units=single,
angle-symbol-over-decimal, retain-explicit-plus]{siunitx}
\usepackage{amsmath}
\usepackage{graphicx}
\usepackage{chemformula}
\usepackage{acro}
\usepackage{layouts}
\usepackage{float}
\usepackage{multirow}
\acsetup{patch/longtable=false}

\received{}

\shorttitle{}
\shortauthors{}

\DeclareSIUnit{\mag}{mag}
\DeclareSIUnit{\pixel}{pixel}
\DeclareSIUnit{\parsec}{pc}
\DeclareSIUnit{\arcsec}{arcsec}
\DeclareSIUnit{\arcmin}{arcmin}
\DeclareSIUnit{\solarlum}{\mbox{$L_\odot$}}
\DeclareSIUnit{\solarmass}{\mbox{$\mathcal{M}_\odot$}}
\DeclareSIUnit{\solarmetal}{\mbox{$Z_\odot$}}
\DeclareSIUnit{\year}{yr}
\DeclareSIUnit{\deg}{deg}
\DeclareSIUnit{\erg}{erg}
\DeclareSIUnit{\dex}{dex}
\DeclareSIUnit{\angstrom}{\textup{\AA}}
\DeclareSIUnit{\radius}{\mbox{$R_{25}$}}
\DeclareSIUnit{\spaxel}{spaxel}

\newcommand{\ha}{H$\alpha$}
\newcommand{\hb}{H$\beta$}
\newcommand{\oiii}{[\ion{O}{3}]}
\newcommand{\nii}{[\ion{N}{2}]}
\newcommand{\oii}{[\ion{O}{2}]}
\newcommand{\sii}{[\ion{S}{2}]}
\newcommand{\siii}{[\ion{S}{3}]}
\newcommand{\hii}{\ion{H}{2}}

\newcommand{\hei}{\ion{He}{1}}

\newcommand{\U}{$\mathcal{U}$}
\newcommand{\logU}{$\log\mathcal{U}$}
\newcommand{\Uoxy}{O$_{32}$}
\newcommand{\Usulf}{S$_{32}$}
\newcommand{\logUoxy}{$\log\text{O}_{32}$}
\newcommand{\logUsulf}{$\log\text{S}_{32}$}

\defcitealias{kewley2019}{K19}
\defcitealias{morisset2016}{M16}
\defcitealias{dors2011}{D11}
\defcitealias{diaz2000}{D00}

\DeclareAcroEnding{possessive}{'s}{'s}

\NewAcroCommand\acg{m}{\acropossessive\UseAcroTemplate{first}{#1}}

\DeclareAcronym{sfr}{short=SFR, long=star formation rate}
\DeclareAcronym{ssfr}{short=sSFR, long=specific star formation rate}
\DeclareAcronym{psf}{short=PSF, long=point spread function}
\DeclareAcronym{agn}{short=AGN, long=active galactic nucleus, long-plural-form=active galactic nuclei}
\DeclareAcronym{lsb}{short=LSB, long=low surface brightness}
\DeclareAcronym{ism}{short=ISM, long=interstellar medium}
\DeclareAcronym{sdss}{short=SDSS, long=Sloan Digital Sky Survey}
\DeclareAcronym{dig}{short=DIG, long=diffuse ionized gas}
\DeclareAcronym{ifs}{short=IFS, long=integral field spectroscopy}
\DeclareAcronym{imf}{short=IMF, long=initial mass function}
\DeclareAcronym{sfh}{short=SFH, long=star formation history, long-plural-form=star formation histories}
\DeclareAcronym{sed}{short=SED, long=spectral energy distribution}
\DeclareAcronym{chaos}{short=CHAOS, long=CHemical Abundances Of Spirals}
\DeclareAcronym{signals}{short=SIGNALS, long={Star-formation, Ionized Gas, and Nebular Abundances Legacy Survey}}
\DeclareAcronym{ccc}{short=CCC, long=concordance correlation coefficient}
\DeclareAcronym{sitelle}{short=SITELLE, long={Spectro-Imageur \`{a} Transform\'{e}e de Fourier pour l'\'{E}tude en Long et en Large des raies d'\'{E}mission}}
\DeclareAcronym{cfht}{short=CFHT, long={Canada-France-Hawaii Telescope}}

\newcolumntype{P}[1]{>{\centering\arraybackslash}p{#1}}

\begin{document}

\title{NGC~628 in SIGNALS: Explaining the Abundance-Ionization Correlation in \hii\ Regions}
\author{Ray Garner, III}
\affiliation{Department of Physics and Astronomy, Texas A\&M University, 578 University Dr., College Station, TX, 77843, USA}
\affiliation{George P.\ and Cynthia W.\ Mitchell Institute for Fundamental Physics \& Astronomy, Texas A\&M University, 578 University Dr., College Station, TX, 77843, USA}

\author{Robert Kennicutt, Jr.}
\affiliation{Department of Physics and Astronomy, Texas A\&M University, 578 University Dr., College Station, TX, 77843, USA}
\affiliation{George P.\ and Cynthia W.\ Mitchell Institute for Fundamental Physics \& Astronomy, Texas A\&M University, 578 University Dr., College Station, TX, 77843, USA}
\affiliation{Department of Astronomy and Steward Observatory, University of Arizona, 933 N.\ Cherry Ave., Tucson, AZ, 85721, USA}

\author{Laurie Rousseau-Nepton}
\affiliation{Dunlap Institute of Astronomy and Astrophysics, University of Toronto, 50 St.\ George St., Toronto, ON, M5S 3H4, Canada}
\affiliation{Department of Astronomy \& Astrophysics, University of Toronto, 50 St.\ George St., Toronto, ON, M5S 3H4, Canada}
\affiliation{Canada-France-Hawaii Telescope, 65-1238 Mamalahoa Hwy, Kamuela, Hawaii 96743, USA}

\author{Grace M.\ Olivier}
\affiliation{Department of Physics and Astronomy, Texas A\&M University, 578 University Dr., College Station, TX, 77843, USA}
\affiliation{George P.\ and Cynthia W.\ Mitchell Institute for Fundamental Physics \& Astronomy, Texas A\&M University, 578 University Dr., College Station, TX, 77843, USA}

\author{David Fern\'{a}ndez-Arenas}
\affiliation{Canada-France-Hawaii Telescope, 65-1238 Mamalahoa Hwy, Kamuela, Hawaii 96743, USA}

\author{Carmelle Robert}
\affiliation{D\'{e}partement de physique, de g\'{e}nie physique et d'optique, Universit\'{e} Laval, Qu\'{e}bec City, QC G1V 0A6, Canada}
\affiliation{Centre de recherche en astrophysique du Qu\'{e}bec, Universit\'{e} de Montr\'{e}al, Montr\'{e}al, QC H3C 3J7, Canada}

\author{Ren\'{e} Pierre Martin}
\affiliation{Department of Physics and Astronomy, University of Hawaii at Hilo, Hilo, HI, 96720, USA}

\author{Philippe Amram}
\affiliation{Aix Marseille University, CNRS, CNES, Laboratoire d'astrophysique de Marseille, 38 Rue Fr\'{e}d\'{e}ric Joliot Curie, Marseille, 13013, France}

\correspondingauthor{Ray Garner}
\email{ray.three.garner@gmail.com}


\begin{abstract}

The variations of oxygen abundance and ionization parameter in \hii\ regions are usually thought to be the dominant factors that produced variations seen in observed emission line spectra. However, if and how these two quantities are physically related is hotly debated in the literature. Using emission line data of NGC~628 observed with SITELLE as part of the \ac{signals}, we use a suite of photoionization models to constrain the abundance and ionization parameters for over 1500 \hii\ regions throughout its disk. We measure an anti-correlation between these two properties, consistent with expectations, although with considerable scatter. Secondary trends with dust extinction and star formation rate surface density potentially explain the large scatter observed. We raise concerns throughout regarding various modeling assumptions and their impact on the observed correlations presented in the literature. 

\end{abstract}

\section{Introduction}

Classical \hii\ regions are large, low-density clouds of partially ionized gas in which star formation has recently taken place. These clouds are ionized by the hot, massive, short-lived stars that emit large amounts of ultraviolet radiation. Despite coming in a wide range of sizes and morphologies, each with its own ionizing populations, one can learn about their chemical compositions and their ionization states by studying their emission line spectra. 

Numerous prescriptions exist in the literature to estimate the metallicity of an \hii\ region\footnote{In this paper ``metallicity'' and ``oxygen abundance'' will be used interchangeably unless otherwise noted.}. The most direct and physically motivated method is to measure the electron temperature, $T_e$, of the ionized gas using the intensity of one or more temperature-sensitive auroral lines like \oiii$\lambda$4363. Unfortunately, these auroral lines are intrinsically faint and not often observed. This is a common issue, so ``strong-line'' calibrations have been presented that use easily measurable strong lines to estimate the oxygen abundance (see \citealt{peimbert2017,perezmontero2017,maiolino2019} for excellent reviews). 

While these strong-line calibrations are considered ubiquitously in the literature, there are still unsolved issues and secondary dependencies remaining. One such dependency is the ionization parameter, \U, defined as the ratio of the number density of incident ionizing photons and the number density of hydrogen atoms \citep{kewley2002}. Thus, the ionization parameter gives an insight into the efficiency of the radiation that ionizes the gas, where the higher the value, the easier the radiation can ionize metals. Unfortunately, this parameter is almost impossible to be measured directly and depends on the internal structure of an \hii\ region. Additionally, those strong-line calibrations derived from photoionization models strongly depend on the input parameters, which are often hard to constrain (e.g., metal depletion, density distribution, geometry, etc.). This might lead to the discrepancy in abundance calibrations where empirical and theoretical calibrations disagree by up to \SI{0.5}{\dex} \citep[e.g.,][]{kewley2008,moustakas2010}. A similar, yet unexplored, situation exists in the calibrations for the ionization parameter, where they differ by an order of magnitude \citep{kreckel2019,mingozzi2020}. Thus, calibrations derived entirely from photoionization models are not to be trusted implicitly and require deeper insight into their effect on observed correlations.

One such correlation potentially impacted by photoionization modeling is between metallicity and the ionization parameter. These are often assumed to be the main contributors to the variations of line ratios among star-forming regions. Early works found an anti-correlation between these two quantities \citep{dopita1986,bresolin1999,dopita2006,maier2006,nagao2006}. \cite{dopita2006} presented a theoretical explanation using a wind-driven bubble model for \hii\ regions where two effects might take place at higher metallicities: (1) stellar winds become more opaque and absorb ionizing photons, reducing the ionization parameter, and (2) the stellar atmospheres scatter photons more effectively, leading to stronger winds, enlarging the \hii\ region and diluting the ionizing flux (see also \citealt{cantiello2007,eldridge2011,xiao2018}). Other effects might also be present, including dust absorption producing a softer radiation field \citep{dopita1986,yeh2012,ali2021}, a correlation between the \ac{imf} and metallicity \citep{martinnavarro2015}, or a correlation between ionization and age \citep{pellegrini2020,scheuermann2023}. Regardless of the cause, other studies have since observationally supported the existence of this anti-correlation \citep[e.g.,][]{perezmontero2014,morisset2016,espinosaponce2022}. 

Interestingly, many studies do not find an anti-correlation between metallicity and the ionization parameter, instead finding either no correlation or a positive correlation \citep[e.g.,][]{kennicutt1996,garnett1997,dors2011,dopita2014,poetrodjojo2018,kreckel2019,mingozzi2020,grasha2022,ji2022}. A search began to find a secondary parameter to explain these positive correlations, or the scatter observed in the anti-correlations. For example, \cite{dopita2014} found a positive correlation between \U\ and  the \ac{sfr} in luminous infrared galaxies, where a higher \ac{sfr} could increase \U\ either due to the higher mass of the ionizing star cluster or a change in the geometry of the cloud. Meanwhile, \cite{pellegrini2020} and \cite{mingozzi2020} found a correlation not between \U\ and \ac{sfr} but between \U\ and \ac{ssfr} as traced by the \ha\ equivalent width. These studies proposed that these relations hold for all galaxies, although other studies have found that the relations are different for individual galaxies \citep{poetrodjojo2018}. 



In this work, we attempt to unravel some of these dependencies by using emission-line fluxes from \hii\ regions in the nearby spiral galaxy NGC~628 collected with the \acl{sitelle} (\acs{sitelle}; \citealt{drissen2019}) as part of the \acl{signals} (\acs{signals}; \citealt{rousseaunepton2019}). NGC~628 is a well-known spiral galaxy seen almost face-on. Its relatively close distance (\mbox{$D = \SI{9.0}{\mega\parsec}$}; \citealt{dhungana2016}) enables its properties to be studied in detail. Notably, given its close distance and subsequently high spatial resolution ($\sim$\SI{35}{\parsec} with \acs{sitelle}; \citealt{rousseaunepton2018}), even the faintest low-luminosity \hii\ regions can be resolved. Thus, we are in a key position to explore the potential correlation between metallicity and ionization parameter in NGC~628. 

While the correlations and anti-correlations found rely on a deep understanding of the physical conditions of \hii\ regions, the uncertainties of photoionization modeling \citep{ji2022} and the resolved scale of observations \citep{kewley2019} still need to be understood. Since we are studying the \hii\ regions of a singular galaxy, we have the ability to tailor photoionization models to our dataset and estimate abundances and ionization parameters in a self-consistent Bayesian framework \citep[e.g.,][]{thomas2018}. This allows for complete freedom in the assumptions and priors used rather than relying on literature calibrations. Our high spatial resolution also gives us unprecedented insight into the role this plays in measuring a (anti-)correlation between these two properties.


This paper is organized as follows. Section~\ref{sec:reduction} describes the observational data and our \hii\ region catalog. In Section~\ref{sec:calib}, we motivate our use of custom photoionization models by exploring the discrepancies between existing ionization parameter calibrations. We describe our \textsc{cloudy} photoionization models and estimate parameters from these models using NebulaBayes in Section~\ref{sec:models}. Given this relatively new technique, we justify its use by recovering numerous gradients and trends already discovered in NGC~628 in Section~\ref{sec:gradients}. Section~\ref{sec:corr} presents the anti-correlation between metallicity and the ionization parameter in NGC~628 and discusses possible secondary parameters causing the large amount of scatter we measure. Finally, we present our conclusions in Section~\ref{sec:conclusion}. 

Throughout this work, we use the following abbreviations for some of the frequently mentioned emission lines. We denote \nii$\lambda$6583, \sii$\lambda\lambda$6717,6731, \siii$\lambda\lambda$9069,9531, \oiii$\lambda\lambda$5007,4959, and \oii$\lambda$3727 as \nii, \sii, \siii, \oiii, and \oii, respectively, unless otherwise noted.

\section{Data Reduction \& \hii\ Region Selection}\label{sec:reduction}

The data for NGC~628 were taken over two observing sessions using the imaging Fourier transform spectrograph \ac{sitelle} at the \ac{cfht} as part of \ac{signals} \citep{rousseaunepton2019}. Full details of our observation and reduction techniques are given in \cite{rousseaunepton2018}. Table~1 of the aforementioned paper gives quantitative information about the filters and total exposure times for our data. Briefly, we summarize our observations here. 

\begin{figure*}
\includegraphics[width=\textwidth,keepaspectratio]{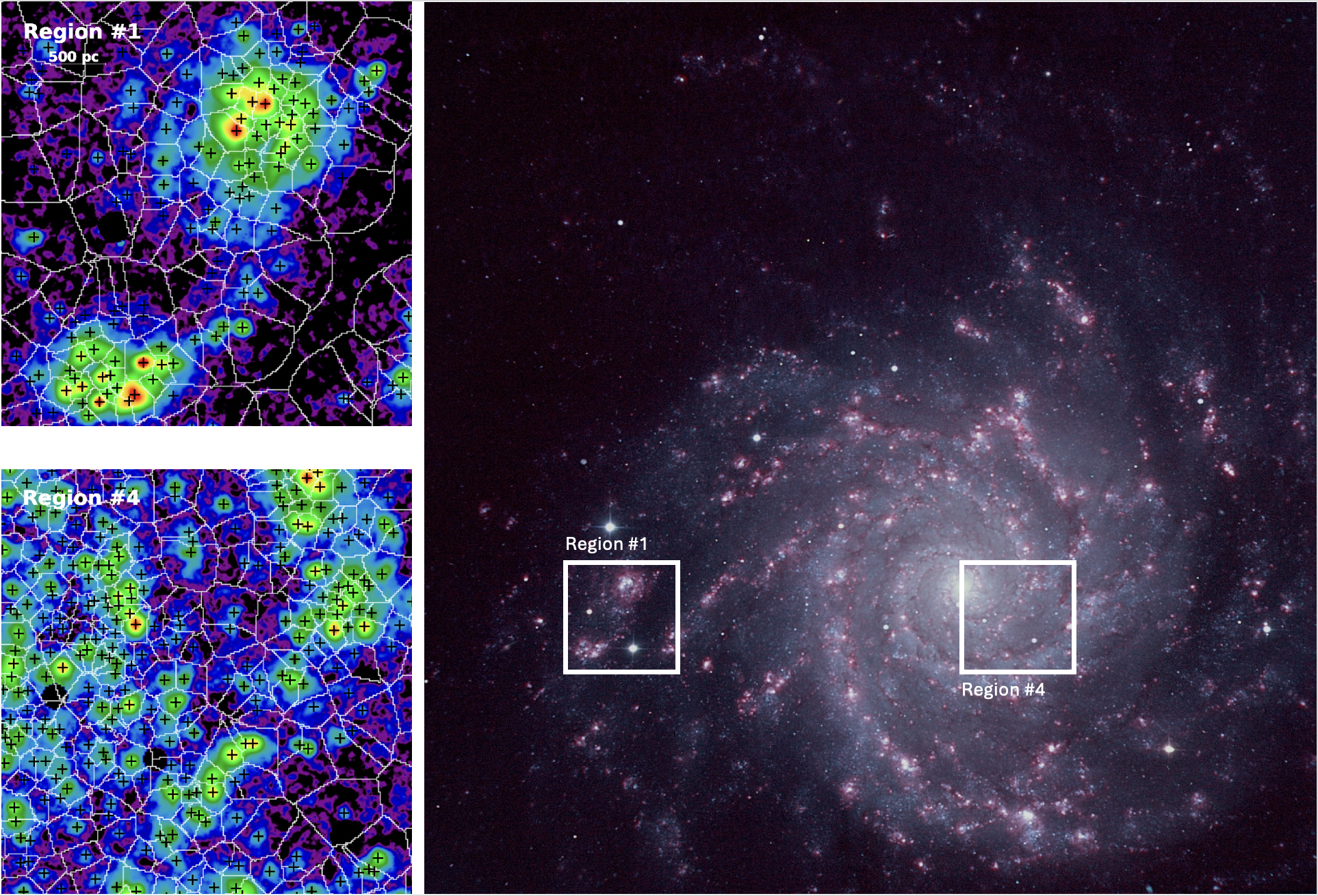}
\caption{Left panels: Regions \#1 (top) and \#4 (bottom) from \cite[][Fig.~7]{rousseaunepton2018}. Each panel shows ionizing sources and their zones of influence drawn over the \ha$+$\hb$+$\oiii\ continuum-subtracted image. The centroid position of each emission peak detected is identified with a cross. The white outlines define the zones of influence surrounding the emission peaks. Each panel measures $\sim$\ang{;;75}$\times$\ang{;;75} (\qtyproduct[product-units=single]{3.25 x 3.25}{\kilo\parsec}) Right panel: the deep image of NGC~628 taken with \acs{sitelle} \citep[][Fig.~2]{rousseaunepton2018}. For each pixel, the three filters were summed together along with the \ha\ intensity map which highlights the emission regions in red on the image. The white boxes indicate the positions of Regions \#1 and \#4. North is up and east is left. The image measures $\ang{;11;} \times \ang{;11;}$ (\qtyproduct[product-units=single]{29 x 29}{\kilo\parsec}).}
\label{image}
\end{figure*}

\ac{sitelle} is a Michelson interferometer inserted into the collimated beam of an astronomical camera system \citep{drissen2019}. Its large FOV ($\ang{;11;} \times \ang{;11;}$) with complete spatial coverage, a high spectral resolution of up to $R \simeq \num{10000}$, and a broad wavelength range from \SIrange{3500}{9000}{\angstrom} with excellent efficiency in the blue part of the spectrum make this an ideal instrument to study the \hii\ regions of any local galaxy. Filters are necessary with \ac{sitelle} to reduce the noise in a selected bandpass. For this project, three filters were used, namely SN1 (\SIrange{3640}{3850}{\angstrom}), SN2 (\SIrange{4840}{5120}{\angstrom}), and SN3 (\SIrange{6480}{6860}{\angstrom}). These filters allow for the measurement of multiple strong emission lines: \oii$\lambda$3727, \hb, \oiii$\lambda\lambda$4959,5007, \nii$\lambda\lambda$6548,6583, \ha, \hei$\lambda$6678, and \sii$\lambda\lambda$6717,6731. The right panel of Figure~\ref{image} shows the deep image produced by adding, for each pixel, the whole signal from the three filters together with an enhanced contribution from the \ha\ intensity map extracted from the line fitting procedure.

The data reduction utilizes the software \textsc{orbs} \citep{martin2015}, which is specifically developed for \ac{sitelle}. \cite{rousseaunepton2018} details the full reduction steps. The extraction software \textsc{orcs} \citep{martin2016}, another software developed for \ac{sitelle}, is used to fit the sinc-shaped line profiles. \cite{rousseaunepton2018} explains how the emission-line regions are detected, giving us flexibility in defining \hii\ regions and their \ac{dig} background. First, the full dataset undergoes sky subtraction, astrometry matching, subtraction of the underlying stellar population, and finally line fitting. The emission peaks are located using a combination of the \ha, \hb, and \oiii\ continuum-subtracted image to build up the signal-to-noise ratio, S/N. These peaks are identified when (1) the pixel intensity is greater than the intensity of at least five immediate surrounding pixels and (2) the total intensity in a $3\times3$ pixel box centered on the emission peak is above the adopted detection threshold fixed by the $3\sigma$ noise level of the flux map. This technique detected a total of \num{4285} emission peaks. The ``zone of influence'', defined by the distance between each pixel and its nearest emission peak, is used to fit a pseudo-Voigt profile where its intensity profile radius determines the size of a region. These region domains are used over the entire data cube. The \ac{dig} is then estimated based on the median intensity for all pixels in an annulus \SI{50}{\parsec} thick centered on the outer limit of a region and subtracted. The left panel of Figure~\ref{image} illustrates this process, showing two regions of NGC~628. The ionizing sources and their zones of influence are drawn over the \ha$+$\hb$+$\oiii continuum subtracted image.

The emission peak catalog of \cite{rousseaunepton2018} consists of RA/Dec coordinates, deprojected radial distances, \ha\ luminosities, \ac{dig} background levels, morphological category, line fluxes and their associated uncertainties. These line fluxes are corrected for extinction assuming a \cite{cardelli1989} extinction curve and Case B recombination \citep{osterbrock2006}. We will be using these reddening-corrected line fluxes unless otherwise noted. 

We here apply data quality cuts to select well-detected \hii\ regions. First, a $\text{S/N} > 5$ is required for all strong emission lines. We utilize the $\text{S/N}_{\text{cross}}$ parameter defined by \cite{rousseaunepton2018}, which takes into account the S/N of two lines simultaneously. Namely, we require that $\text{S/N}_{\text{cross}} > 5$ for \oiii/\hb, \oii/\hb, \nii/\ha, and \sii/\ha. This requirement excludes about \SI{55}{\percent} of the regions. 

\begin{figure*}
\includegraphics[width=\textwidth,keepaspectratio]{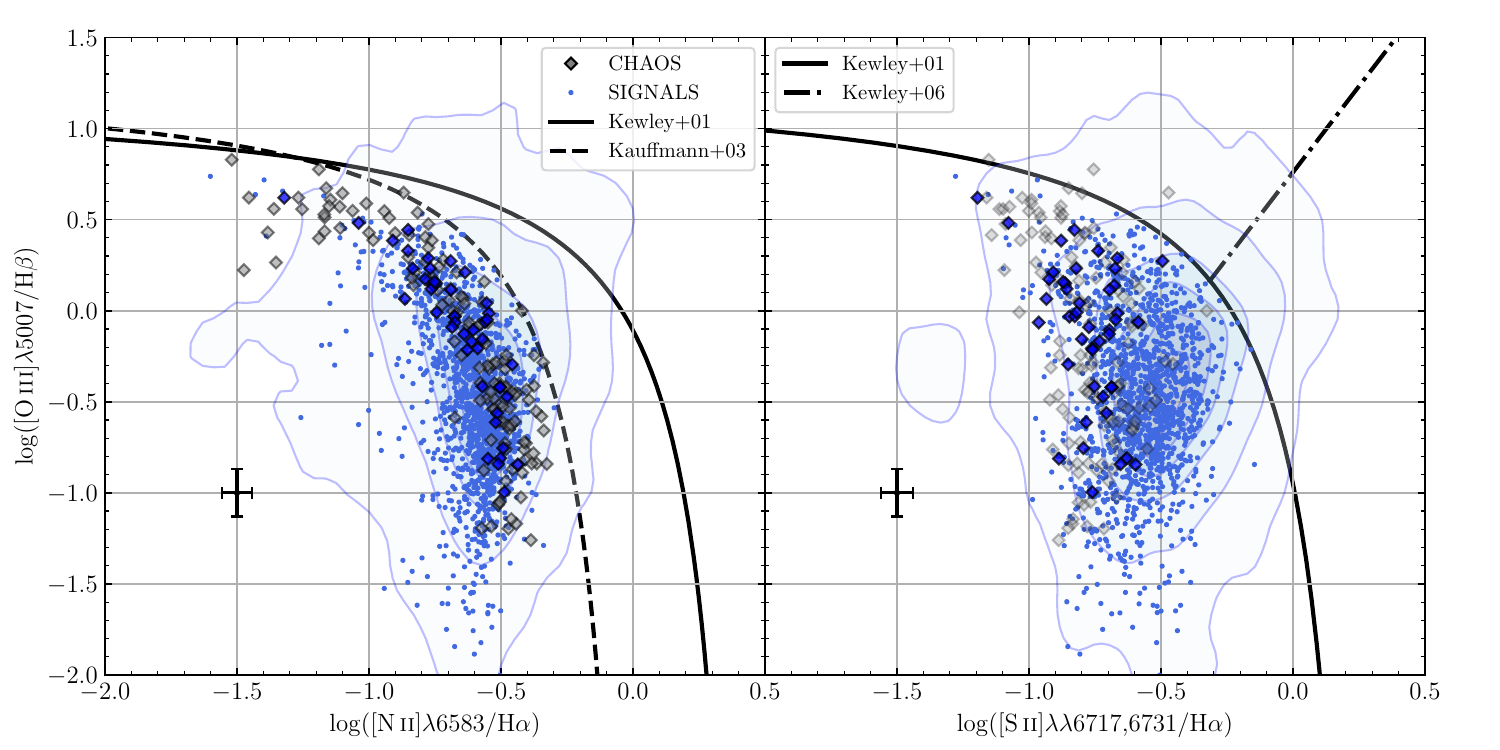}
\caption{The \nii\ (left) and \sii\ (right) BPT \citep{baldwin1981,veilleux1987} diagrams for the \ac{signals} and \acs{chaos} samples. The \ac{signals} data for NGC~628 before any data cuts are shown by the blue contours enclosing \SI{99}{\percent}, \SI{90}{\percent}, \SI{70}{\percent}, \SI{50}{\percent}, \SI{30}{\percent}, and \SI{10}{\percent} from lightest to darkest blue. The remaining \hii\ regions after data cuts are shown by the light blue points with characteristic error bars shown to the left in each plot. The \acs{chaos} sample are the grey diamonds with those for NGC~628 in dark blue; their uncertainties are smaller than the size of the data point. In the left panel, the dashed line is the pure star-formation demarcation from \cite{kauffmann2003}. In both panels, the solid line is the extreme starburst demarcation from \cite{kewley2001}. In the right panel, dash-dot line is the Seyfert/LINER line from \cite{kewley2006}.}
\label{bpt-obs}
\end{figure*}

Next, we apply a cut on the BPT \citep{baldwin1981,veilleux1987} diagrams to ensure the emission lines are consistent with photoionization by massive stars. In the \oiii/\hb\ versus \nii/\ha\ diagram, we require the regions to be consistent with the more stringent \cite{kauffmann2003} empirical line. In the \oiii/\hb\ versus \sii/\ha\ diagram, the regions must be consistent with the \cite{kewley2001} line. This excludes a further $\sim$\SI{3}{\percent}. 

We require that the uncertainty in the color excess, $E(B-V)$, is less than five times the value of the color excess. We trim those outliers with extremely high extinction values, requiring $E(B-V) < 0.8$ or $A_V < 2.5$. This excludes another $\sim$\SI{6}{\percent} of the regions. Finally, we require that all emission lines are detected. This final cut excludes only a further five regions. In total, our \hii\ region catalog consists of \num{1532} \hii\ regions.

In addition to our \ac{signals} data for NGC~628, we also include four galaxies from the \ac{chaos} project, namely NGC~628, M51, M101, and NGC~3184 (\citealt{berg2015,croxall2015,croxall2016,berg2020}, respectively) totaling \num{175} \hii\ regions. Using the Multi-Object Double Spectrograph on the Large Binocular Telescope, the \ac{chaos} project  is a spectroscopic survey of bright \hii\ regions in nearby spiral galaxies. Their continuous and wide spectral coverage (\SIrange{3200}{10000}{\angstrom}), as well as auroral line-based abundance estimates, serves as a useful comparison sample and will be helpful in setting Bayesian priors in Section~\ref{sub:nebulabayes}.

Figure~\ref{bpt-obs} shows the \nii\ and \sii\ BPT diagrams for our \ac{signals} dataset and the \ac{chaos} dataset. After the data quality cuts described above, our \ac{signals} data for NGC~628 fall in the region of the plots consistent with photoionization by massive stars. As mentioned and performed by \cite{rousseaunepton2018}, comparing our sample to the \ac{chaos} sample not only demonstrates our ability to reproduce line ratios measured by spectroscopic surveys but also includes objects with a larger range of physical properties (masses, ages, stellar content, \ac{ism} properties, etc.). 

\section{A Problem with Ionization Parameter Calibrations?}\label{sec:calib}

As mentioned in the Introduction, numerous calibrations exist in the literature to estimate the oxygen abundances of \hii\ regions. Detailed analysis of these calibrations has revealed a systematic discrepancy between them, leading to abundance differences of up to \SI{0.5}{\dex} \citep[e.g.,][]{kewley2008,moustakas2010}. There appears to be a similar, yet unexplored, issue with the calibrations for the ionization parameter. Several studies have mentioned that existing calibrations differ by an order of magnitude \citep{kreckel2019,mingozzi2020}. However, aside from the pioneering work of \cite{ji2022}, who determined the strength of different factors leading to the differences between publicly available photoionization models, the discrepancies between the calibrations have yet to receive equal attention.

Briefly, let us begin with some useful definitions related to the ionization state of the gas in an \hii\ region.\footnote{For more details, the interested reader is referred to the foundational textbooks of \cite{osterbrock2006} and \cite{draine2011}.} Ionic species in \hii\ regions are in ionization equilibrium with the terms dependent on the properties of the ionizing source and the gas gathered together to define the ionization parameter: 
\begin{equation}\label{eq:U-strom}
	\mathcal{U} = \frac{Q_0}{4\pi R_S^2 cn_e}.
\end{equation}
Here, $Q_0$ is the production rate of hydrogen-ionizing photons, $n_e$ is the electron density, and $c$ is the speed of light. This is known as the ``dimensionless'' ionization parameter where the ``dimensionful'' parameter is usually defined as $q = \mathcal{U}c$ in \si{\centi\metre\per\second}. Here we have made the \cite{stromgren1939} approximation, using the size of a Str\"{o}mgren sphere, $R_S$, to define \U. The definition of the Str\"{o}mgren radius is based on the balance between ionization and recombination rates assuming Case B recombination \citep{osterbrock2006}, where
\begin{equation}
	R_S = \left(\frac{3Q_0}{4\pi\alpha_B\varepsilon n_{\text{H}}^2}\right)^{1/3} \approx \left(\frac{3Q_0}{4\pi\alpha_B\varepsilon n_{e}^2}\right)^{1/3}.
\end{equation}
In the above expression, we assume all of the gas is ionized, so the hydrogen density and electron density are approximately equal, $n_{\text{H}} \approx n_e$. We also introduce the Case B recombination coefficient, $\alpha_B$, and the volume-filling factor of the gas, $\varepsilon$, defined assuming that the gas is structured in clumps that are surrounded by a lower-density medium, where it is unity for a homogeneous constant-density gas and decreases as the density of clumps increases \citep{kennicutt1984}. For simplicity, we can assume \mbox{$\alpha_B = \num{2.56e-13}T_4^{-0.83} \, \si{\centi\metre\cubed\per\second}$}, where $\mathbf{T_4 = T_e/\SI{e4}{\kelvin}}$ \citep{draine2011}. This simplifies the ionization parameter to: 
\begin{equation}
	\mathcal{U} \propto Q_0^{1/3}n_e^{1/3}\varepsilon^{2/3}T_4^{-0.55}.
\end{equation}
We see that the ionization parameter has a weak dependence on the ionizing photon production and gas density, and is somewhat more dependent on the filling factor and electron temperature. 

However, there are several issues with this common definition of the ionization parameter. While the Str\"{o}mgren sphere assumption is convenient, it is probably not physically motivated as it assumes a sphere of constant density that surrounds the ionizing source. However, in real \hii\ regions, feedback from stellar winds carves out a cavity around the ionizing source, making the ionized gas a shell rather than a sphere. Real \hii\ regions also show various complicated substructures and geometries. The Str\"{o}mgren sphere also assumes that the \hii\ region is radiation-bounded where no ionizing photons escape instead of density-bounded. 

The main problem is that while it carries useful information about the ionizing source and gas geometry, the ionization parameter is not directly observable. The ionization parameter is instead often estimated using measurements of emission line ratios that have a sensitivity to the ionization state of the gas, pairs of low and high ionization states of the same element, in conjunction with predictions from photoionization models. For instance, the ratios \oiii$\lambda\lambda$5007,4959/\oii$\lambda$3727 or \siii$\lambda\lambda$9069,9531/\sii$\lambda\lambda$6717,6731, hereafter \Uoxy\ and \Usulf, respectively, are commonly used since these ratios are good indicators for \ch{O^{2+}}/\ch{O^+} and \ch{S^{2+}}/\ch{S^+} \citep{aller1942,diaz2000,kewley2002,kewley2019}.

Since each photoionization model used to calibrate observed line ratios and \U\ depends on the assumptions inherent to each model, this produces offsets between different calibrations. We investigate these offsets and differences in the \Uoxy\ and \Usulf\ calibrations using data from \ac{signals} and \ac{chaos}, respectively. Namely, we use the calibrations of \cite{kewley2019}, \cite{morisset2016}, \cite{dors2011}, and \cite{diaz2000} (hereafter \citetalias{kewley2019}, \citetalias{morisset2016}, \citetalias{dors2011}, and \citetalias{diaz2000}, respectively), all of whom have published both \Uoxy\ and \Usulf\ calibrations. In the case of the calibrations of \citetalias{kewley2019} where there is a published abundance dependence, we utilized their \nii/\oii\ abundance calibration as it has no dependence on the ionization parameter (\citealt{kewley2002}; \citetalias{kewley2019}). \citetalias{morisset2016} has published \Uoxy\ calibrations using two different geometries, a thin shell and a filled sphere; we utilize both. Finally, we compare these calibrations to the assumptions of a Str\"{o}mgren sphere, outlined above. We determine $Q_0$ empirically from the dust-corrected \ha\ luminosity assuming the relationship of \cite{osterbrock2006} and no escape or dust absorption of ionizing photons, i.e.,
\begin{equation}
	Q_0 \, [\si{\per\second}] = \num{7.35e11}L_{\mathrm{H}\alpha} \, [\si{\erg\per\second}].
\end{equation}
To measure the electron density, we used PyNeb \citep{luridiana2015}, considering the ratio of \sii$\lambda$6717/\sii$\lambda$6731 and assuming $T_e = \SI{e4}{\kelvin}$. We fix the volume filling factor $\varepsilon = 0.01$, a reasonable assumption for \hii\ regions \citep[e.g.,][]{kennicutt1984,cedres2013}. 

\begin{figure*}
\includegraphics[width=\textwidth,keepaspectratio]{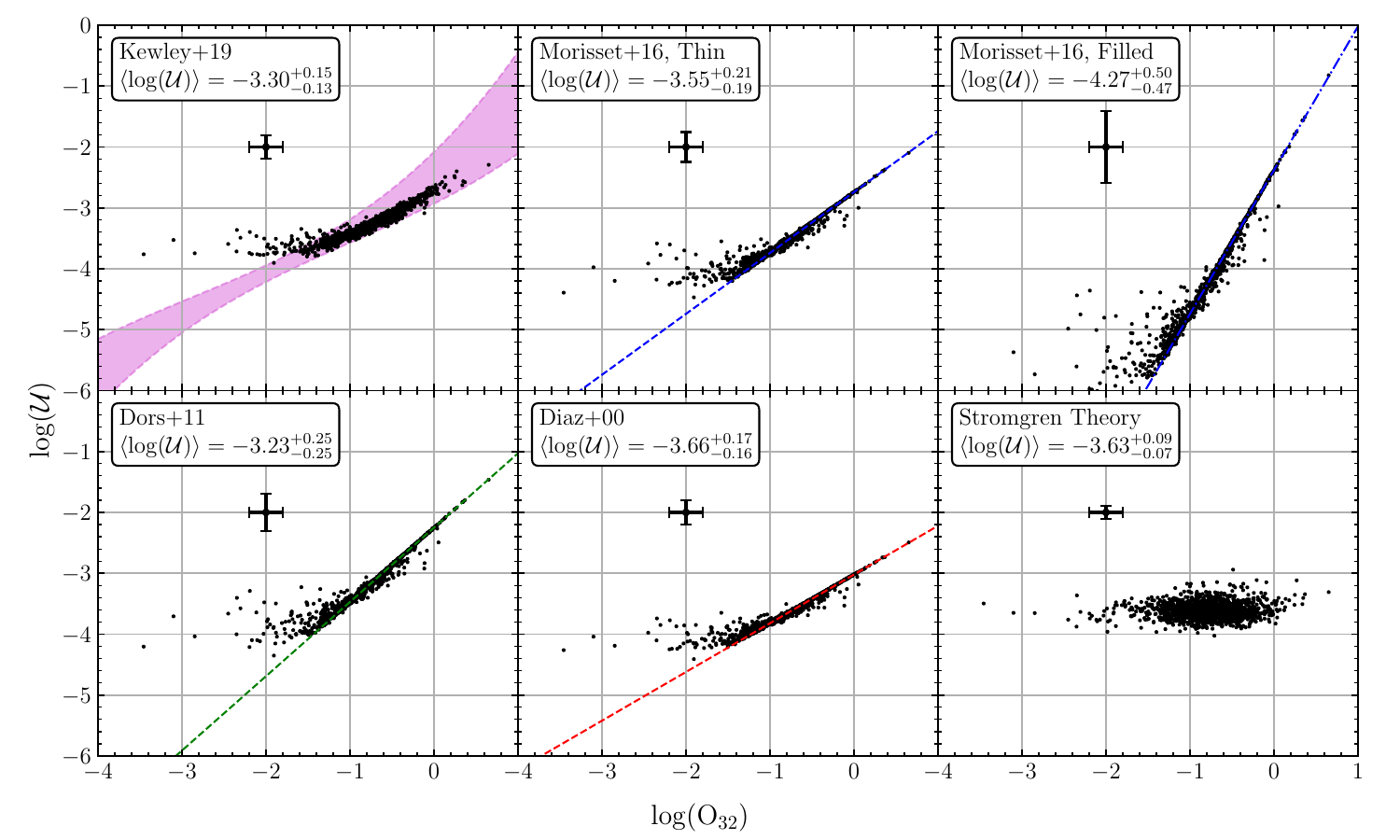} \\
\includegraphics[width=\textwidth,keepaspectratio]{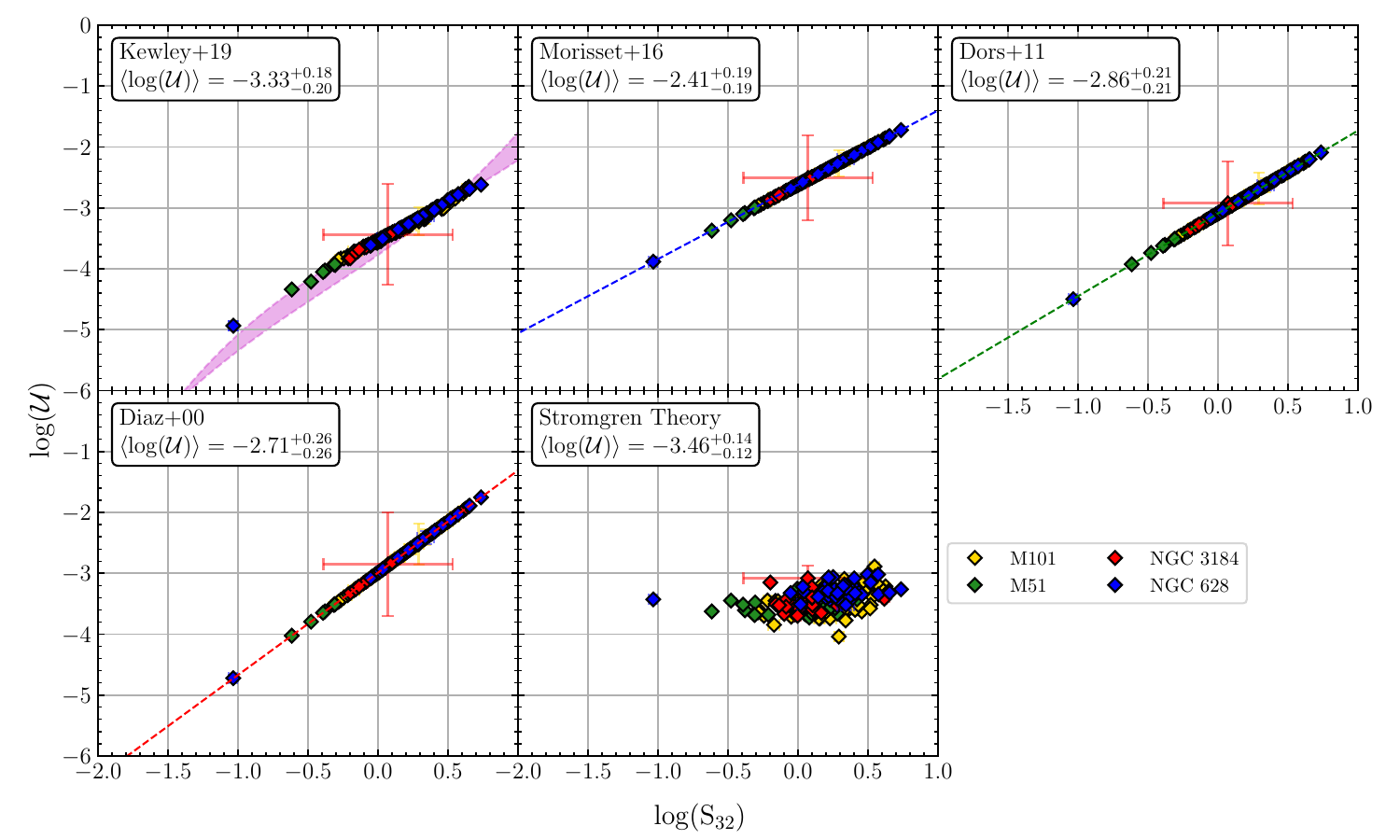}
\caption{\logU\ calibrations from the literature. The top group of six panels shows the \logUoxy-\logU\ calibrations applied to the \ac{signals} data. Characteristics error bars are shown in the upper left. The bottom group of five panels shows the \logUsulf-\logU\ calibrations applied to the \ac{chaos} data. Colors indicating different \ac{chaos} galaxies are indicated in the legend. Individual error bars are shown, often smaller than the size of the data point. Each panel shows a different calibration given in the legend as well as the median and quartiles for each calibration.}
\label{calib-lit}
\end{figure*}

We used a Monte Carlo method to apply these calibrations while including the observational uncertainties associated with the emission-line fluxes by running the calculations for 500 trials assuming Gaussian uncertainties. In this way, we take the median of these 500 trials as the estimate of \U\ and the standard deviation as the uncertainty on \U. Figure~\ref{calib-lit} shows the \logUoxy-\logU\ calibrations from the literature applied to the \ac{signals} dataset on the top, and the \logUsulf-\logU\ calibrations applied to the \ac{chaos} dataset on the bottom. Reported in each plot are the median and quartiles for each calibration.

Looking first at the \logUoxy-\logU\ calibrations in Figure~\ref{calib-lit}, we see that the data at low \logUoxy\ flares away from the calibration lines in each case. These data have very large uncertainties in \logUoxy, upwards of \SI{2}{\dex} caused by large uncertainties in \oiii, which results in large uncertainties on the \logU\ values, $\sim$\SI{0.6}{\dex}, due to the random Monte Carlo sampling. However, most of our data points lie along the calibration line. The median values of \logU\ for the calibrations are approximately the same at $\log \mathcal{U} \simeq -3.5$ with \citetalias{kewley2019} and \citetalias{dors2011} predicting slightly higher values. Interestingly, the \citetalias{morisset2016} filled sphere calibration predicts much lower values of $\log U \simeq -4.3$, which disagrees strongly with the Str\"{o}mgren sphere estimation. Much more concerning is that none of the calibrations exactly agree with one another. The disagreement is likely caused by the \Uoxy\ ratio's strong dependence on the metallicity as well as \ac{ism} pressure in metal-rich galaxies (\citealt{kewley2002}; \citetalias{kewley2019}) since only some models control for these dependencies. 

The \logUsulf-\logU\ calibrations do not fare much better. This is especially concerning given that all of the \ac{chaos} \hii\ regions are the bright, massive, and young \hii\ regions easily accessible by most surveys. The \Usulf\ ratio is frequently claimed to be a better estimation of the ionization parameter than \Uoxy\ as it lacks a dependency on metallicity. However, comparing the median values between \logUoxy\ and \logUsulf\ calibrations shows a higher scatter between \logUsulf-\logU\ calibrations than between \logUoxy-\logU\ calibrations. The historical difficulty photoionization codes have had reproducing the strengths of the far-red \siii\ lines might cause the higher scatter. This difficulty was a result of a lack of an accurate estimate of the dielectronic recombination coefficient for the transition $\text{S}^{2+} \rightarrow \text{S}^+$ \citep{izotov2009,belfiore2022}. Recent updates to the photoionization code \textsc{cloudy} \citep{chatzikos2023} have fixed this, resulting in an increase in the predicted flux of the \sii\ lines and a decrease in \Usulf\ by \SIrange{20}{50}{\percent} at a fixed \Uoxy\ \citep{badnell2015}. These changes could explain why the \logUsulf-\logU\ calibration of \citetalias{kewley2019} predicts a lower \logU\ more in line with the \logUoxy-\logU\ calibrations than the others.

\begin{deluxetable*}{lc}
\tablecaption{Input Parameters for Literature Photoionization Models \label{calib}}
\tablehead{\colhead{Parameter} & \colhead{Values}}
\startdata
\multicolumn{2}{c}{K19 model \citep{kewley2019}} \\
Photoionization Code & \textsc{mappings} v5.1 \citep{sutherland2018} \\
\logU & \numlist[list-final-separator={, }]{-4;-3.75;-3.5;-3.25;-3;-2.75;-2.5;-2.25;-2} \\
$\log(Z/Z_\odot)$ & \numlist[list-final-separator={, }]{-1.06;-0.46;-0.16;0.24;0.54} \\
$\log(n_{\text{H}}/\si{\per\cubic\centi\metre})$ & 1 \\
Geometry & Plane-parallel \\
Stellar \ac{sed} & \textsc{starburst99} v7 \citep{leitherer2014} \\
Solar Abundance Set & \cite{nicholls2017} \\
Nitrogen Prescription & \cite{nicholls2017} \\
Dust Depletion Factor & \cite{jenkins2009} \\ \hline 
\multicolumn{2}{c}{M16 model \citep{morisset2016}} \\
Photoionization Code & \textsc{cloudy} v13.03 \citep{ferland2013} \\
\logU & \numlist[list-final-separator={, }]{-4;-3.75;-3.5;-3.25;-3;-2.75;-2.5;-2.25;-2;-1.75;-1.5} \\
$\log(Z/Z_\odot)$ & \numrange[range-phrase={ to }]{-0.53}{0} \\
$\log(n_{\text{H}}/\si{\per\cubic\centi\metre})$ & 1 \\
Geometry & Plane-parallel \& Spherical \\
Stellar \ac{sed} & POPSTAR \citep{molla2009} \\
Solar Abundance Set & \cite{asplund2009} \\
Nitrogen Prescription &  $\log(\mathrm{N/O})$: \numlist[list-final-separator={, }]{-1.5;-0.75;0;0.75;1.5} \\
Dust Depletion Factor & Default depletion set in \textsc{cloudy} \citep{cowie1986,jenkins1987} \\ \hline 
\multicolumn{2}{c}{D11 model \citep{dors2011}} \\
Photoionization Code & \textsc{cloudy} v8.00 (last described in \citealt{ferland1998})  \\
\logU & \numlist[list-final-separator={, }]{-3.5;-3;-2.5;-2;-1.5} \\
$\log(Z/Z_\odot)$ & \numlist[list-final-separator={, }]{-1.3;-0.7;-0.4;-0.2;0;0.3} \\
$\log(n_{\text{H}}/\si{\per\cubic\centi\metre})$ & 2.3 \\
Geometry & Plane-parallel \\
Stellar \ac{sed} & \textsc{starburst99} v6 \citep{leitherer2010} \\
Solar Abundance Set & \cite{allendeprieto2001} \\
Nitrogen Prescription & \cite{vilacostas1993} \\
Dust Depletion Factor & \cite{garnett1995} \\ \hline 
\multicolumn{2}{c}{D00 calibration \citep{diaz2000} using the models of \cite{diaz1991}} \\
Photoionization Code & \textsc{cloudy} v74\tablenotemark{a} \citep{ferland1989} \\
\logU & \numlist[list-final-separator={, }]{-4;-3.5;-3;-2.5} \\
$\log(Z/Z_\odot)$ & \numlist[list-final-separator={, }]{0;0.3} \\
$\log(n_{\text{H}}/\si{\per\cubic\centi\metre})$ & 1 \\
Geometry & Spherical \\
Stellar \ac{sed} & Single-star models of \cite{mihalas1972} \\
Solar Abundance Set & \cite{stasinska1990} \\
Nitrogen Prescription & Unspecified \\
Dust Depletion Factor & Unspecified \\ \hline
\enddata
\tablenotetext{a}{\cite{diaz1991} did not specify which version of \textsc{cloudy} they used. However, it should be no later than v74 according to the code release information at \url{https://gitlab.nublado.org/cloudy/cloudy/-/wikis/CloudyOld}.}
\end{deluxetable*}

Table~\ref{calib} shows that each calibration is derived from photoionization models with different built-in assumptions and modeling codes. \cite{ji2022} performed an extensive analysis on a different set of publicly available photoionization models. They found three major factors contributing to the differences among those models: updated atomic data (mentioned above), the stellar \ac{sed}, and the gas-phase chemical abundance set. Broadly speaking, we find the same important differences in these models. We refer the interested reader to the excellent discussion of \cite{ji2022} but briefly mention a few key points particular to these models. 

The stellar \ac{sed} determines the relative number of ionizing photons and helps set the emission-line ratios. Most of the models use well-established and recent \textsc{starburst99} or PopStar \acp{sed}, which, at least for the timescales necessary for \hii\ regions, produce similar numbers of ionizing photons between them \citep{molla2009}. The lone exception is the calibration of \citetalias{diaz2000}, which uses the models of \cite{diaz1991}. These models utilized single-star models with atmospheres taken from \cite{mihalas1972}. Aside from this not being physically correct for \hii\ regions consisting of ionizing star clusters of different ages, earlier work showed that these stellar atmospheres are unsuitable for modeling \hii\ regions due to their lack of treatment for heavy element bound-free opacity \citep{borsenberger1982,evans1991}. Despite these strong cautions, the \citetalias{diaz2000} calibration does show a remarkable agreement with the others considered here. Perhaps another model parameter offsets the issues caused by these softer \acp{sed}. 

The chemical abundances are fundamental for setting the line ratios emitted by the \hii\ regions. While four separate solar abundance sets are used, these are further modified by the chosen nitrogen prescription, i.e., the relation between the N/O ratio and metallicity, and the dust depletion factors. The chemical abundance sets and nitrogen prescriptions of each of the four models in Table~\ref{calib} are approximately the same: the nitrogen prescriptions of \cite{nicholls2017} and \cite{vilacostas1993} used by \citetalias{kewley2019} and \citetalias{dors2011}, respectively, differ only in the low-metallicity regime by about \SI{0.2}{\dex}. Meanwhile \citetalias{morisset2016} uses the N/O ratio as an input parameter, and it is unclear what, if any, nitrogen prescription the models of \citetalias{diaz2000} used. 

The chosen dust depletion factors also has an important effect on the emitted spectrum of an \hii\ region. Dust depletion has two effects: reducing the intensities of the lines emitted by depleted elements, and strengthening the lines emitted by non-depleted elements as the removed coolants raise the equilibrium temperature. All of the dust depletion models deplete oxygen by varying degrees, the \cite{jenkins2009} depletion factors used by \citetalias{kewley2019} model depletes nitrogen, and none deplete sulfur\footnote{It remains an open question whether or not sulfur depletes in the \ac{ism}. See \cite{jenkins2009} for a discussion.}. While this would suggest that the dust depletion should not affect the \logUsulf-\logU\ calibrations, removing other elemental coolants will strengthen the remaining emission lines making the model spectrum appear more highly ionized. Perhaps this explains some of the differences observed in the \logUsulf-\logU\ calibrations, while the differences observed in the \logUoxy-\logU\ calibrations are more intertwined with other model factors.

In summary, the atomic data, the stellar \ac{sed}, and the final gas-phase chemical abundance set are likely responsible for the differences between these models and their resulting ionization parameter calibrations. Unfortunately, without publicly available model grids or detailed descriptions of the model inputs, further analysis makes it difficult to say which, if any, of these three input parameters is most responsible. As \cite{ji2022} mention, there is also the possibility of degeneracies between these three inputs: a harder stellar \ac{sed} might cancel out the underestimation of elemental abundances or overestimation of depletion factors. Therefore, using these published calibrations is ill-advised without understanding the assumptions inherent in their models. 


\section{Photoionization Models and Parameter Estimation}\label{sec:models}


Additionally, the most popular line ratio used to estimate \logU\ and the one directly available to us, \Uoxy, has a secondary dependence on metallicity \citep{kewley2002}. Since we wish to explore the correlation between oxygen abundance and ionization parameter, instead of adopting any particular calibrator for each, we choose to estimate \U\ and $12 + \log(\text{O/H})$ for each of our \hii\ regions in a self-consistent way, i.e., through our photoionization modeling. This way, we can avoid mixing the uncertainties between two calibration assumptions, one for \U\ and another for $12 + \log(\text{O/H})$. The details of our photoionization models and our estimation of individual \hii\ regions' parameters are described below.

\subsection{\textsc{cloudy} Models}

We used the photoionization code \textsc{cloudy} v23.01 \citep{chatzikos2023}. The input \ac{sed} for the models were computed using the code \textsc{starburst99} \citep{leitherer2014}. We assume a \cite{kroupa2001} \ac{imf}, and a fixed total mass of \SI{e7}{\solarmass} over \SI{100}{\mega\year}, selecting the \ac{sed} at \SI{2.5}{\mega\year} as the \ac{sed} for our \hii\ regions. From previous studies, NGC~628 has \hii\ regions with a gas-phase abundance of approximately solar \citep[e.g.][]{mccall1985,ferguson1998,sanchez2011,zou2011,berg2015}. Making the reasonable assumption of a tight link between the gas-phase abundance and stellar-phase abundance of the ionizing stars, i.e., recently formed OB stars, we only utilized \acp{sed} with solar stellar abundances, $Z_\odot = 0.02$. When computing these \acp{sed}, we used the \cite{pauldrach2001} and \cite{hillier1998} stellar atmospheres and the high mass-loss Geneva evolutionary tracks \citep{meynet1994}. 

We set the hydrogen density of the ionized gas cloud to \SI{100}{\per\cubic\centi\metre}, which is the median electron density calculated from the ratio of the \sii\ lines, and assume an electron temperature of \SI{e4}{\kelvin}. The gas pressure is considered to be constant throughout the cloud. We include dust grains with typical \ac{ism} abundance, which we scale with the metallicity of the cloud. Metal depletion onto the dust grains is computed using the values given by \cite{cowie1986} and \cite{jenkins1987}. The solar abundance we use is taken from \cite{grevesse2010} with $12 + \log(\mathrm{O/H})_\odot = 8.69$. We scale all other elements with oxygen except nitrogen, carbon, and sulfur. We let the nitrogen abundance be a free parameter in our models while we adopt the C/O prescription of \cite{dopita2013}. We refit these data using a simple quadratic function:
\begin{equation}
	\text{C/H} = 0.249 \times (\text{O/H})^2 + 3.515 \times \text{O/H} + 5.322.
\end{equation}
For sulfur, we adopt the S/O prescription of \cite{berg2020} who found a constant $\log(\text{S/O}) = \num{-1.34 \pm 0.15}$. This value differs from the solar value by about \SI{0.23}{\dex} \citep{grevesse2010}. 

\begin{figure}
\includegraphics[width=\columnwidth,keepaspectratio]{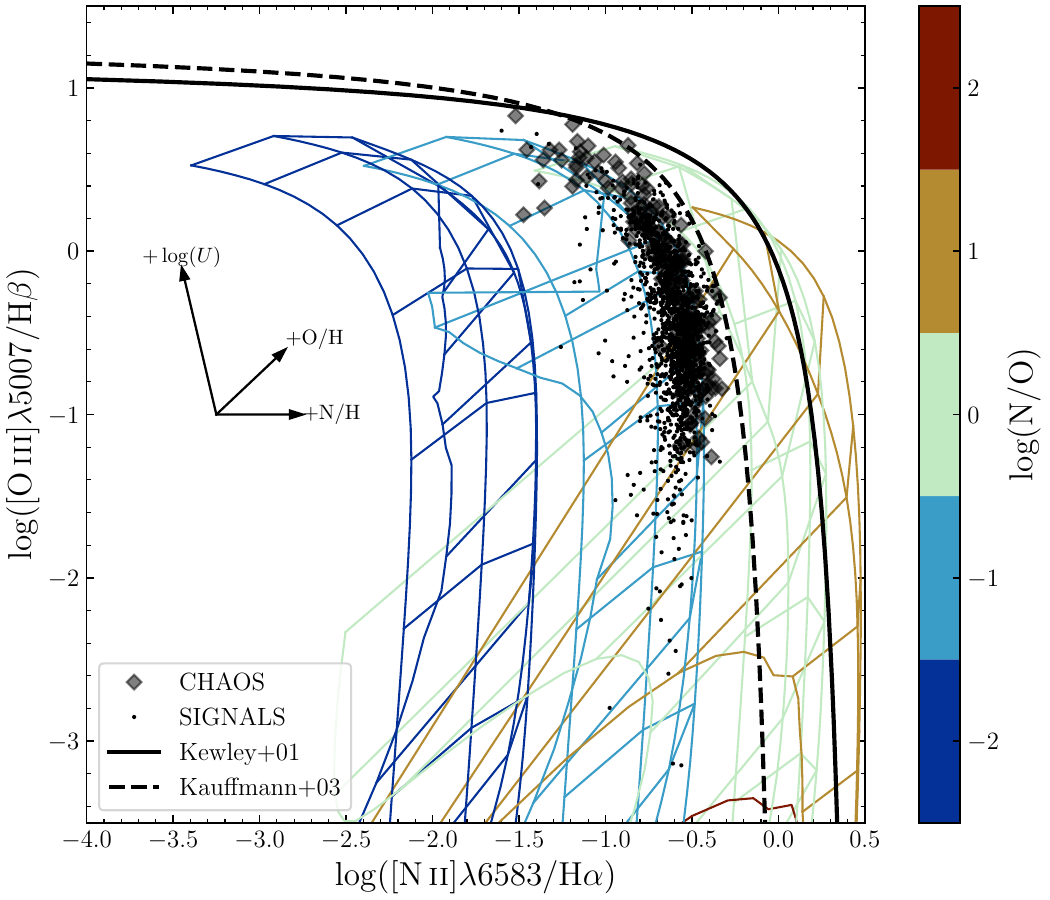}
\caption{The \nii\ BPT diagram for the \ac{signals} (black dots) and \ac{chaos} (grey diamonds) samples with a subset of our photoionization models overlayed. The models are colored by their N/O ratios. Arrows are given to indicate in which direction \logU, O/H, and N/H increase. Demarcation lines from \cite{kauffmann2003} and \cite{kewley2001} are shown as dotted and solid black lines, respectively.}
\label{bpt-models}
\end{figure}

To set up a grid, we vary the cloud's ionization parameter, oxygen abundance, and nitrogen abundance. The range we adopt for each are \mbox{$-5 \leq \log \mathcal{U} \leq -1.5$} in \SI{0.1}{\dex} steps, \mbox{$-4.5 \leq \log\mathrm{O/H} \leq -2.5$} in \SI{0.1}{\dex} steps, and \mbox{$-6.5 \leq \log\mathrm{N/H} \leq -2.5$} in \SI{0.25}{\dex} steps. This results in \num{12852} total models. Figure~\ref{bpt-models} shows a subset of our models on the \nii\ BPT diagram and how they overlap with both the \ac{signals} and \ac{chaos} data. Evident from this figure is that our models cover the emission-line space occupied by our data well, except for a few very high ionization regions. Additionally, we can easily read off the plot that most of our data (\SI{78}{\percent}) has $-1 \lesssim \log(\mathrm{N/O}) \lesssim 0$, which is consistent with the \ac{chaos} N/O ratios measured using direct abundances \citep{berg2020}. 

\subsection{Extracting Properties with NebulaBayes}\label{sub:nebulabayes}

Given the relatively large parameter space outlined above, we need a method that efficiently estimates the parameters for each \hii\ region in our sample. We use the Bayesian analysis code NebulaBayes \citep{thomas2018}, chosen as it is a generalization of two previous Bayesian codes IZI \citep{blanc2015} and \textsc{bond} \citep{valeasari2016}. While it comes with a pre-made model grid for \hii\ regions \citep{thomas2018}, NebulaBayes is entirely customizable, allowing for custom model grids, error weighting, choice of priors, and choice of lines that influence the likelihood distribution, among others. Given this information and observed emission-line fluxes, NebulaBayes returns an inferred value and uncertainty for each property varied in the model grid. This code has been used extensively in estimating \ac{agn} properties \citep[e.g.,][]{thomas2019,radovich2019,zovaro2020,perezdiaz2021,polimera2022,peluso2023,li2024} but little work has used it to estimate the properties of \hii\ regions \citep[e.g.,][]{espinosaponce2022,li2024}. As such, an extensive investigation into the proper choice of NebulaBayes parameters is necessary. 

Here, we investigate three NebulaBayes parameters: error weighting, choice of lines that influence the likelihood distribution, and choice of priors. In order to compare our model properties with observed properties, we use the \ac{chaos} dataset, which has emission-line fluxes, including the \siii\ lines, and measured oxygen and nitrogen abundances. However, it is important to remember that the measured abundances come from electron temperature measurements, which are offset by as much as \SI{0.5}{\dex} from photoionization grid abundances \citep[e.g.,][]{moustakas2010}. 

First, we pick how the errors on the observed line fluxes are weighted. Normally, one would simply use the observed line flux uncertainties as the weights, but due to the changing resolution across the three \ac{sitelle} filters ($R \simeq 600$ in SN1 and SN2 and $R \simeq 1800$ in SN3; \citealt{rousseaunepton2018}), the observed uncertainties change with wavelength as well. The median \oii\ and \oiii\ uncertainties in the \ac{signals} dataset is $\sim$\SI{30}{\percent}, while that of \sii\ and \nii\ is $\sim$\SI{15}{\percent} and $\sim$\SI{10}{\percent}, respectively, despite the weaker line flux. Compare this to \ac{chaos} for which the uncertainties are approximately uniform across wavelength at $\sim$\SI{1}{\percent}---as such, using just the observed uncertainties as the weights would shift the NebulaBayes solution towards models that match the \nii\ and \sii\ lines at the expense of the oxygen lines. Therefore, we choose to adopt for all of the \ac{sitelle} lines a \SI{10}{\percent} error as the input observational errors for NebulaBayes to match our most well-determined emission lines.

Second, we choose which observed emission lines influence the likelihood distribution. We chose to output our modeled emission lines as fractions of \hb, and since we have already corrected our data for extinction, the hydrogen lines \ha\ and \hb\ do not provide any additional constraining power. We do not include these lines in the likelihoods. We must include the \nii, \oii, and \oiii\ emission lines since ratios between these lines, i.e., \nii/\oii, $R_{23} = (\text{\oiii} + \text{\oii})/\text{\hb}$, and $\text{O}_{32} = \text{\oiii}/\text{\oii}$, are proxies for N/H, O/H, and \U, respectively \citep[e.g.,][]{pagel1979,kewley2002,kobulnicky2004}. 

Ultimately, we decided against including the sulfur lines in the likelihoods for a few reasons. First, \sii\ comes from the outskirts of an \hii\ region, which does not coincide with the production sites of other strong lines (e.g., see Fig.~1 and discussion of \citealt{mannucci2021}). Second, any density structure in the \hii\ region will change the strength of \sii\ with respect to the other lines, and our \textsc{cloudy} models assume a constant electron density throughout the entirety of the modeled \hii\ region. Similarly, while we usually assume that the S/O ratio is constant \citep{garnett1997,izotov2006,berg2020}, recent work has questioned that assumption, finding S/O decreases with increasing O/H \citep{vilchez1988,dors2016,diaz2022}, perhaps due to different production sites of sulfur and oxygen \citep{goswami2024}. Additionally, there are known problems with matching the \sii\ and \siii\ fluxes with photoionization models. The \sii\ lines are generally weaker by $\sim$\SI{0.1}{\dex} than observed \citep{levesque2010,dopita2013,mingozzi2020}. The discrepancy between observed and modeled \siii\ fluxes has been well-reported in the literature \citep[e.g.,][]{dinerstein1986,garnett1989,ali1991}, and is likely caused by limitations in modeling stellar atmospheres and/or in the atomic data for sulfur \citep{garnett1989,badnell2015}. 


\begin{figure*}
\includegraphics[width=\textwidth,keepaspectratio]{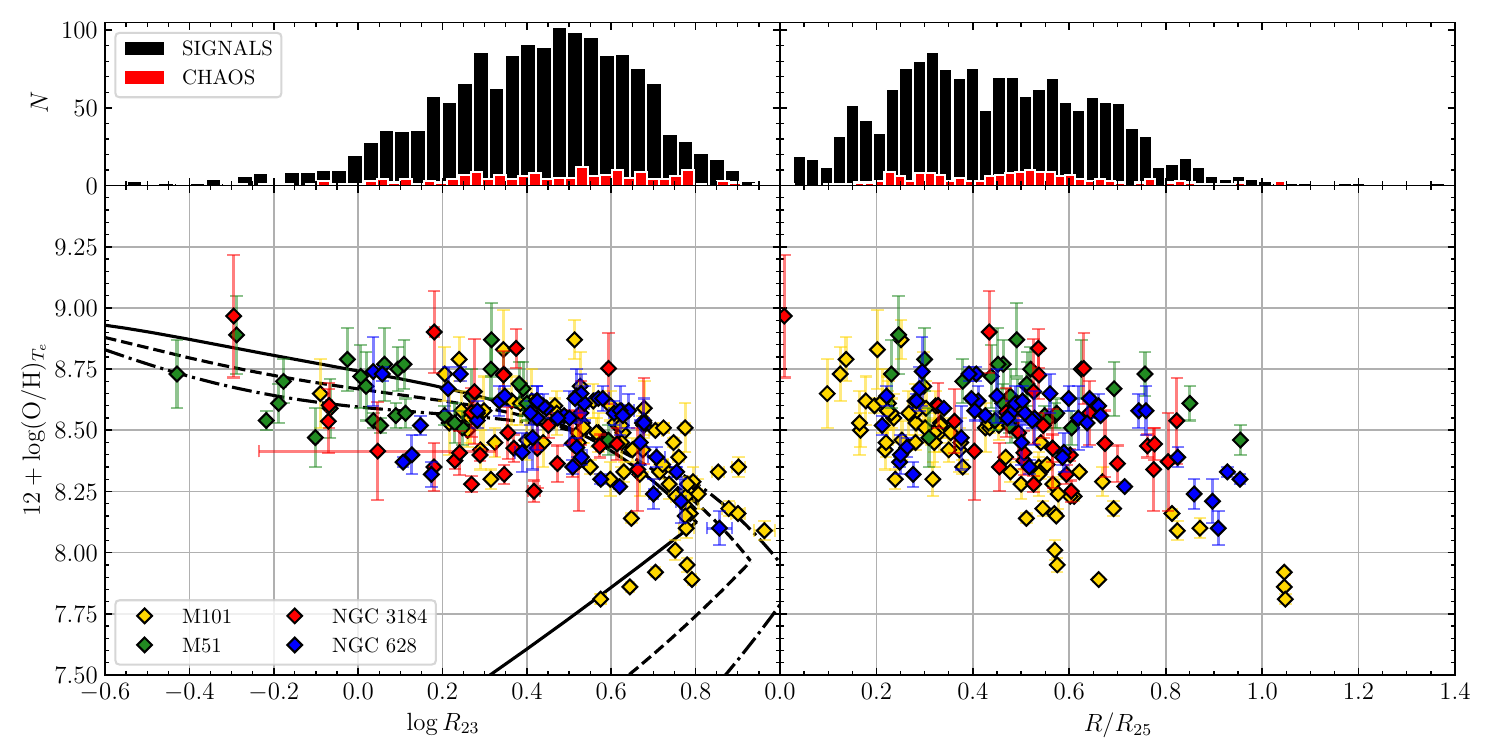}
\caption{Left: the $R_{23}$-O/H relation for the \ac{chaos} data. Included is the $R_{23}$ prescription of \cite{kobulnicky2004} for comparison, shifted down by \SI{0.5}{\dex} to lie on the \ac{chaos} data. The different lines represent different \logU: $-4$ (solid), $-3$ (dashed), and $-2$ (dash-dot). Right: the oxygen abundance gradients for the \ac{chaos} data. In both panels, the colors are as in Figure~\ref{calib-lit}. At the top of both panels are histograms comparing the $R_{23}$ and $R_{25}$ distributions of the \ac{chaos} (red) and \ac{signals} (black) data.}
\label{chaos-r23}
\end{figure*}

Finally, we select the priors. NebulaBayes is very flexible in the shape of the priors, and we choose to use the ``line-ratio prior'' feature. In this case, NebulaBayes calculates a prior over the entire model grid based on the ratio of a pair of lines. We implement a custom ``line-sum prior'' for the $R_{23}$. Numerous studies have used some combination of these or other line ratios as priors \citep[e.g.,][]{valeasari2016,thomas2018,mingozzi2020,li2024}.


Since we are utilizing several strong line ratios, in particular $R_{23}$ which is well-known to be double-valued with oxygen abundance, this set of priors will return double-valued abundance solutions. In order to break this degeneracy, we must rely on assumptions about how the oxygen abundance behaves in the galaxy. Using the \ac{chaos} data as a comparison in Figure~\ref{chaos-r23}, the metal-rich branch of the $R_{23}$-O/H relation is preferred up to large radii, $R < 0.8R_{25}$. Data outside this radial range would prefer the metal-poor branch, but this radial range accounts for only $\sim$\SI{5}{\percent} of our \ac{signals} data for NGC~628, so we can safely assume that most of the \hii\ regions in our sample will lie on the upper, metal-rich branch. Therefore, we convolve an additional Gaussian prior on the oxygen abundance, $\mathcal{N}(9,0.5)$, to select the metal-rich solution from NebulaBayes. We stress that this additional prior does not force the solutions to have oxygen abundances of 9 but simply skews the posteriors to favor the metal-rich solution. A similar technique, albeit done on a region-to-region basis, was used by \cite{strom2018} on their sample of high-redshift galaxies. 


\begin{figure*}
\includegraphics[width=\textwidth,keepaspectratio]{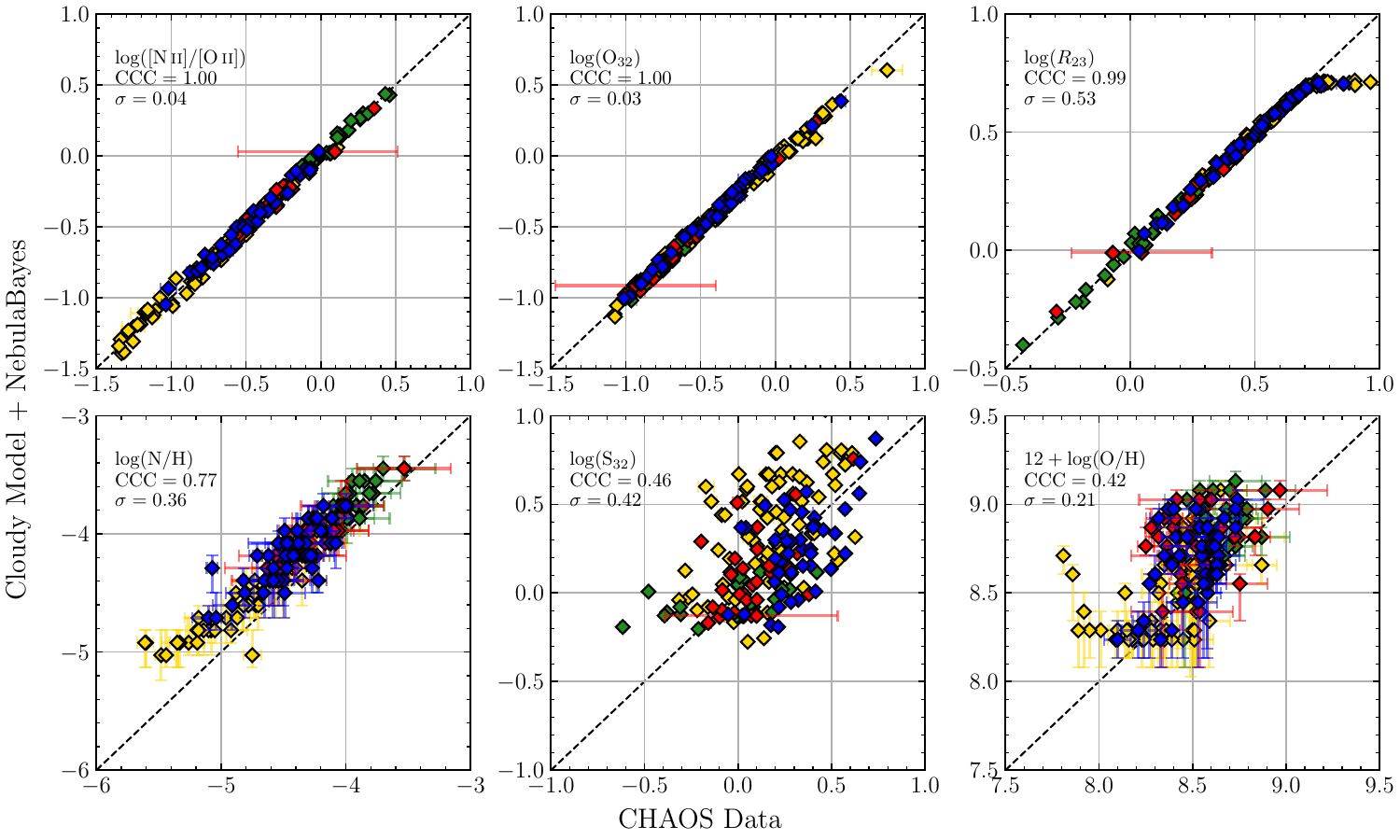}
\caption{Each panel compares the values of one property between that measured by \ac{chaos} on the $x$-axis and our model grid through NebulaBayes on the $y$-axis. The properties compared are labeled inside each panel. Colors are the same as in Figure~\ref{calib-lit}: M101 is gold, M51 is green, NGC~3184 is red, and NGC~628 is blue. In all panels, error bars are shown for the \ac{chaos} data, while NebulaBayes only returns uncertainties for the nitrogen and oxygen abundances. The \acs{ccc} (\acl{ccc}; \citealt{lin1989}), which measures the deviation of a relationship from the $1:1$ line (where $+1$ is a perfect agreement and $-1$ is a perfect disagreement) is reported as well as the scatter, $\sigma$.}
\label{chaos-compare}
\end{figure*}

Figure~\ref{chaos-compare} shows how the resulting derived line ratios and physical properties compare to those obtained by \ac{chaos}. The \nii/\oii\ and \Uoxy\ ratios agree very well, as does the $R_{23}$ up to about $\log(R_{23}) \sim 0.7$. We are likely seeing the effect of forcing these regions to lie on the upper, metal-rich branch of the $R_{23}$-O/H relation since at these $R_{23}$ values, they would no longer be metal-rich. Most of these regions are in M101, but those few in NGC~628 lie at radii greater than $R > 0.8R_{25}$. Again, this only accounts for \SI{5}{\percent} of our \ac{signals} data, so it is likely that most of the galaxy lies on the metal-rich branch. Meanwhile, the line ratio \Usulf\ broadly agrees between our models and \ac{chaos}, although with a considerably large scatter. This is expected since we have removed the sulfur lines' constraining power from our analysis. The nitrogen abundance agrees well within the uncertainties except at very low nitrogen abundance. Finally, we find that the oxygen abundance broadly agrees, though our models show a distinct offset towards higher abundances. This is expected since this comparison involves electron temperature abundances and photoionization model abundances, which are known to be discrepant \citep{kewley2008,moustakas2010}. 

Given the above analysis, and especially the comparisons in Figure~\ref{chaos-compare}, we are confident that we can use NebulaBayes to estimate reasonable physical properties of \hii\ regions. In the next section, we apply NebulaBayes to the \ac{signals} dataset for NGC~628 and investigate the bulk properties and correlations of its \hii\ regions. 

\section{Recovering Gradients in NGC~628}\label{sec:gradients}

Using the priors, lines, and observed line errors described in the previous section, we apply the \mbox{NebulaBayes} algorithm to our \ac{signals} dataset.  We also convolve the posterior distributions of the nitrogen and oxygen abundances to estimate the N/O ratio for our dataset.


In the case of the N/O ratio, 63 (\SI{4}{\percent}) \hii\ regions have unbound posteriors and are treated as upper limits. These regions are unbound in only N/O and are bound in the other three parameters. However, they are at the upper ends of the nitrogen and oxygen abundance distributions, so it is likely that these regions have nonzero N/O ratios but are simply badly fit by a combination of the model grid and NebulaBayes. Calculating the N/O ratios for this set of regions by subtracting the nitrogen and oxygen abundances gives N/O ratios of approximately the solar value, bringing them in line with the rest of the distribution. Therefore, we choose to include these regions with unbound N/O posteriors in our analysis. 

The interquartile ranges in the model parameters for this subsample, including those with upper limits on N/O, are
\begin{align*}
	12 + \log(\mathrm{O/H}) &= [8.61, 8.92], \\
	\log \mathcal{U} &= [-3.07, -2.79],  \\
	\log(\mathrm{N/H}) &= [-4.39, -4.08], \text{ and} \\
	\log(\mathrm{N/O}) &= [-0.84, -0.37].
\end{align*}
These characteristic ranges are consistent with ranges from inferences of these parameters by other means \citep{rosalesortega2011,sanchez2015,espinosaponce2022} as discussed next.

\begin{figure*}
\includegraphics[width=\textwidth,keepaspectratio]{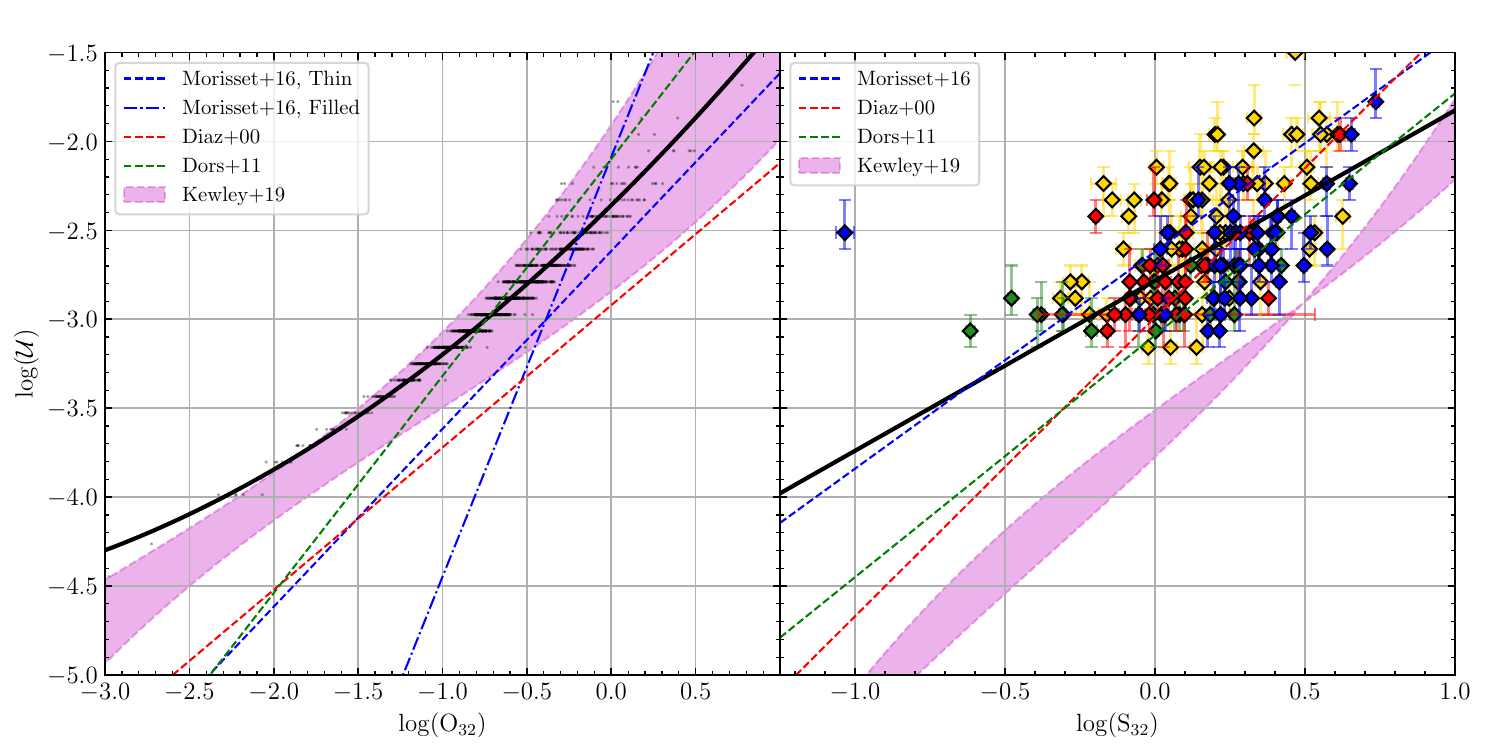}
\caption{Left: the \Uoxy-\logU\ calibration for the \hii\ regions in NGC~628 for \ac{signals}. Individual \hii\ regions are shown as small black points which are in \SI{0.1}{\dex} bins in \logU. The solid black line is the fit to the \ac{signals} data, while the colored lines and bands are calibrations from the literature. Right: the \Usulf-\logU\ calibration of the \hii\ regions in \ac{chaos}. Colors are the same as in Figure~\ref{calib-lit}. The solid black line is the fit to the \ac{chaos} data while the colored lines and bands are calibrations from the literature. See the text for more details.}
\label{logU-data-calib}
\end{figure*}

Figure~\ref{logU-data-calib} shows the \Uoxy-\logU\ and \Usulf-\logU\ calibrations in the left and right panels, respectively. In the left panel, we use a quadratic function to fit individual \hii\ regions from the \ac{signals} dataset. In the right panel, we fit \hii\ regions from the \ac{chaos} dataset using a linear function. Both functional fits and their scatters and \cite{spearman1904} correlation coefficients are given in Table~\ref{fits}. Also shown in both panels are the calibrations from Section~\ref{sec:calib}. Most of the linear calibrations only match our data at high \Uoxy\ with approximately the same slope. In contrast, the filled sphere prescription of \citetalias{morisset2016} does not match our data at all. The calibration of \citetalias{kewley2019} agrees best with nearly all of our \hii\ regions lying within their band of models with a small deviation at low \logU. 

\begin{deluxetable*}{lclccccccc}
\tablecaption{Fit to Gradients and Calibrations \label{fits}}
\tablehead{\colhead{$y$} & \colhead{$x$} & \colhead{Equation} & \colhead{$\sigma_{\text{raw}}$} & \colhead{$\sigma_{\text{bin}}$} & \colhead{$\rho_{\text{raw}}$} & \colhead{$p_{\text{raw}}$} & \colhead{$\rho_{\text{bin}}$} & \colhead{$p_{\text{bin}}$}}
\startdata
\logU\ & \logUoxy\ & $\begin{aligned} y &= +(\num{0.10 \pm 0.01})x^2 + (\num{0.93 \pm 0.02})x \\ &\qquad - (\num{2.36 \pm 0.01}) \end{aligned}$ & 0.07 & & +0.98 & \num{4.9e-22} & & \\
\logU\ & \logUsulf\ & $y = +(\num{0.96 \pm 0.09})x - (\num{2.78 \pm 0.02})$ & 0.29 & & +0.53 & \num{4.8e-14} & & \\
$12 + \log(\mathrm{O/H})$ & $R/R_{25}$ & $y = -(\num{0.65 \pm 0.06})x + (\num{9.04 \pm 0.05})$ & 0.15 & 0.09 & $-0.75$ & \num{1.3e-35} & $-0.93$ & \num{5.8e-7} \\
\logU\ & $R/R_{25}$ & $y = +(\num{0.53 \pm 0.14})x - (\num{3.20 \pm 0.11})$ & 0.30 & 0.16 & $+0.31$ & \num{2.4e-35} & $+0.63$ & \num{1.2e-2} \\
$\log(\mathrm{N/O})$ & $R/R_{25}$ & $y = -(\num{1.07 \pm 0.09})x - (\num{0.18 \pm 0.07})$ & 0.23 & 0.13 & $-0.78$ & \num{1.6e-35} & $-0.95$ & \num{9.5e-8} \\
\logU\tablenotemark{a} & $12 + \log(\mathrm{O/H})$ & $y = -(\num{0.50 \pm 0.09})x + (\num{1.50 \pm 0.76})$ & 0.30 & 0.12 & $-0.21$ & \num{6.8e-17} & $-0.83$ & \num{3.7e-6} \\
\logU\tablenotemark{b} & $12 + \log(\mathrm{O/H})$ & $y = -(\num{0.67 \pm 0.08})x + (\num{3.05 \pm 0.68})$ & 0.31 & 0.09 & $-0.46$ & \num{2.7e-40} & $-0.94$ & \num{4.5e-9} \\
$12 + \log(\mathrm{O/H})$\tablenotemark{a} & $E(B-V)$ & $y = +(\num{0.11 \pm 0.05})x + (\num{8.75 \pm 0.02})$ & 0.22 & 0.08 & $+0.11$ & \num{2.2e-5} & $+0.53$ & \num{5.4e-2} \\
$12 + \log(\mathrm{O/H})$\tablenotemark{b} & $E(B-V)$ & $y = +(\num{0.19 \pm 0.07})x + (\num{8.76 \pm 0.03})$ & 0.22 & 0.06 & $+0.22$ & \num{3.3e-9} & $+0.55$ & \num{4.3e-2} \\
\logU\tablenotemark{a} & $E(B-V)$ & $y = -(\num{0.61 \pm 0.08})x - (\num{2.76 \pm 0.03})$ & 0.30 & 0.06 & $-0.30$ & \num{2.4e-32} & $-0.95$ & \num{6.1e-8} \\
\logU\tablenotemark{b} & $E(B-V)$ & $y = -(\num{0.65 \pm 0.11})x - (\num{2.70 \pm 0.05})$ & 0.30 & 0.08 & $-0.38$ & \num{7.7e-27} & $-0.86$ & \num{3.8e-5} \\
$12 + \log(\mathrm{O/H})$\tablenotemark{a} & $\log\Sigma_{\text{H}\alpha}$ & $y = +(\num{0.02 \pm 0.02})x + (\num{8.15 \pm 0.65})$ & 0.22 & 0.09 & $+0.13$ & \num{8.4e-7} & $+0.28$ & \num{2.8e-1} \\
$12 + \log(\mathrm{O/H})$\tablenotemark{b} & $\log\Sigma_{\text{H}\alpha}$ & $y = -(\num{0.01 \pm 0.02})x + (\num{9.30 \pm 0.66})$ & 0.22 & 0.07 & $+0.05$ & \num{1.8e-1} & $-0.15$ & \num{6.3e-1} \\
\logU\tablenotemark{a} & $\log\Sigma_{\text{H}\alpha}$ & $y = +(\num{0.01 \pm 0.03})x - (\num{3.42 \pm 1.01})$ & 0.31 & 0.11 & $+0.09$ & \num{2.8e-4} & $+0.13$ & \num{6.2e-1} \\
\logU\tablenotemark{b} & $\log\Sigma_{\text{H}\alpha}$ & $y = +(\num{0.04 \pm 0.03})x - (\num{4.59 \pm 1.00})$ & 0.30 & 0.11 & $+0.03$ & \num{3.6e-1} & $+0.41$ & \num{1.7e-1} \\
\enddata
\tablecomments{Fits to various relations in this paper. The $y$ and $x$ variables are given in the first two columns. The resulting best fit is given in Column 3 with uncertainties on all coefficients. Columns 4 and 5 list the residual scatter in the raw unbinned data, $\sigma_{\text{raw}}$, and in the binned data, $\sigma_{\text{bin}}$. Columns 6 and 7 list the Spearman correlation coefficient, $\rho$, and corresponding $p$-value for the raw unbinned data, while Columns 8 and 9 list the same for the binned data.}
\tablenotetext{a}{Fitted to all \hii\ regions.}
\tablenotetext{b}{Fitted to only \hii\ regions with \ha\ $\text{S/N} \geq 15$.}
\end{deluxetable*}

The right panel of Figure~\ref{logU-data-calib} presents the \Usulf-\logU\ calibration for the \ac{chaos} dataset since this dataset has the \siii\ lines. In stark contrast with the left panel, we see that only the calibration presented by \citetalias{kewley2019} disagrees with the data. It is unknown what causes this disagreement, although it could be related to some of the model assumptions discussed in Section~\ref{sec:calib}, particularly the atomic data. What is more apparent is the large amount of scatter seen in the \ac{chaos} data, whereas other studies have generally shown a tight correlation with little scatter between \Usulf\ and \logU\ (\citealt{diaz1991}; \citealt{kewley2002}; \citetalias{dors2011}; \citetalias{morisset2016}). However, the large scatter observed here is likely due to our NebulaBayes algorithm not using the sulfur lines to constrain \logU. Adding the sulfur lines into NebulaBayes would require including a harder prior on the oxygen abundances, namely $\mathcal{N}(9,0.1)$, making direct comparisons to \ac{signals} difficult.  

\begin{figure}
\includegraphics[width=\columnwidth,keepaspectratio]{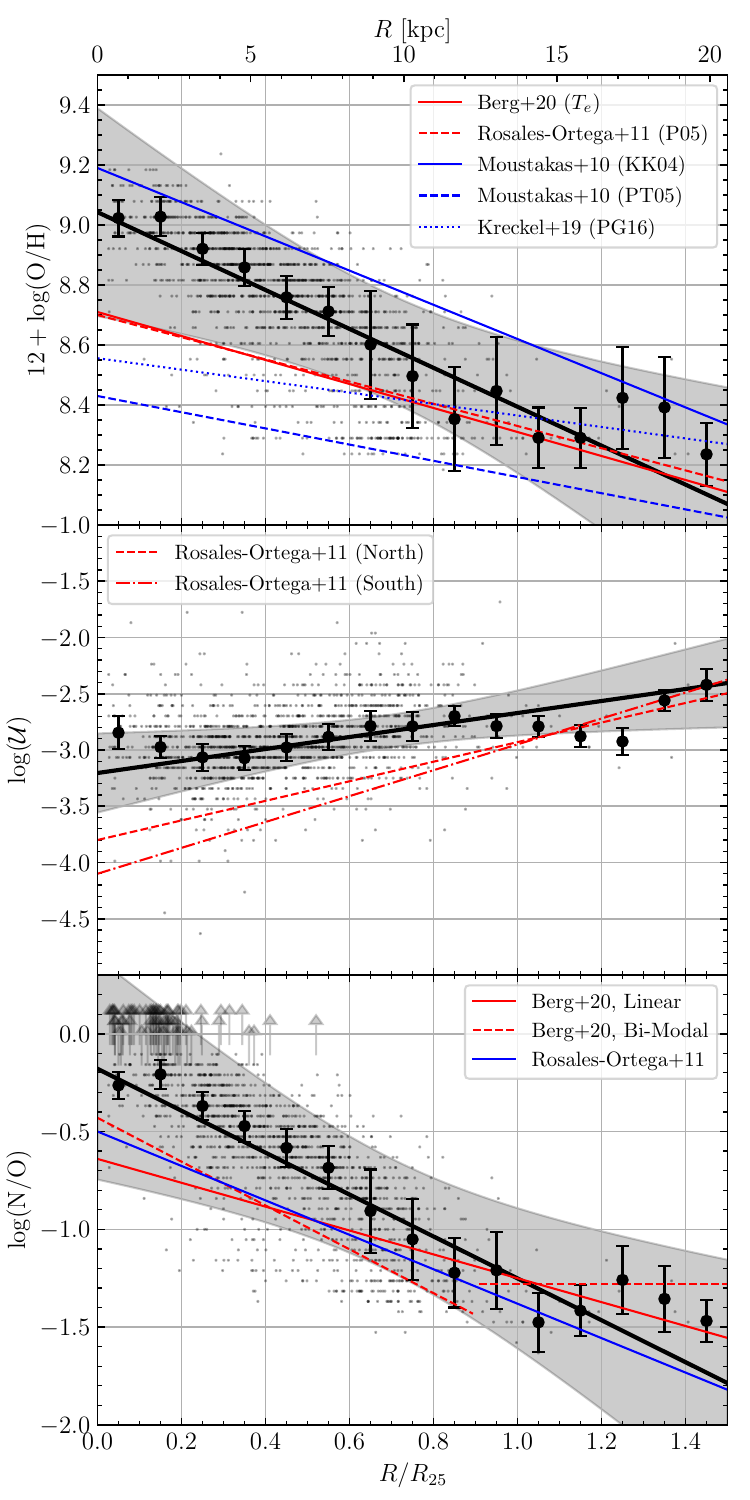}
\caption{Physical parameter radial gradients in NGC~628. In all panels, individual \hii\ regions are small black points which are binned in radius of width $0.1R_{25}$ as the large black points with error bars. The solid black line is a fit to the binned data with the shaded region being the $1\sigma$ uncertainty to the fit. Fitted gradients from the literature are also shown as the colored lines indicated in the legends. Top panel: the oxygen abundance gradient. In the legend, the first citation is the study while the second citation abbreviated in parentheses is the calibration method used: P05 \citep{pilyugin2005_ff}, KK04 \citep{kobulnicky2004}, PT05 \citep{pilyugin2005}, PG16 \citep{pilyugin2016}. Middle panel: the \logU\ radial gradient. Bottom panel: the N/O radial gradient. Those \hii\ regions with only lower limits are shown as faint upwards arrows. These are not included in the fit.}
\label{gradients}
\end{figure}

Figure~\ref{gradients} shows the recovered radial gradients for three parameters: oxygen abundance (top panel), ionization parameter (middle panel), and N/O ratio (bottom panel). In order to fit a radial gradient, we take the median in bins with a width of $0.1R_{25}$. We derive uncertainties on the bins using a Monte Carlo algorithm, which randomly samples each data point within its uncertainties \num{1000} times. The uncertainty on the bin is then the standard deviation of these samples. The gradient is then fit to the binned data. Table~\ref{fits} reports the resulting fits, the scatter in the raw and binned data, and the Spearman correlation coefficients for both the raw and binned data. In what follows, we compare our gradients with those from the literature, noting differences and similarities. 

Starting with the oxygen abundance gradient in the top panel of Figure~\ref{gradients}, we see that our fitted gradient closely matches the slope of that reported by \cite{moustakas2010} using the theoretical $R_{23}$ calibration of \cite{kobulnicky2004} (solid blue line). This agreement is not surprising as we have given NebulaBayes the $R_{23}$ ratio as a prior. Our gradients are the same within the uncertainties with the \cite{moustakas2010} gradient shifted upwards by $\sim$\SI{0.2}{\dex}. We have a steeper gradient than those estimated by electron temperature measurements \citep{rosalesortega2011,berg2020} (red dashed and solid lines, respectively) with ours shifted upwards by $\sim$\SI{0.3}{\dex}. Similarly, the gradient reported by \cite{moustakas2010} using the empirical $P$-method of \cite{pilyugin2005} (dashed blue line) shows a much lower and shallower gradient, shifted downwards from the $T_e$ gradients by a further $\sim$\SI{0.3}{\dex}. These are known issues in comparing abundances derived with different calibrations, and we refer the reader to the discussion of \cite{moustakas2010}. Despite these concerns, the fact that the abundances estimated through NebulaBayes broadly agree with those from more established methods is comforting. 


Turning to the ionization parameter gradient in the middle panel of Figure~\ref{gradients}, we recover a weak positive gradient. We note that it is not the two binned points beyond $1.3R_{25}$ that is determining the positive gradient; rather it is being determined by the points between $0.3R_{25}$ and $0.9R_{25}$. Most galaxies show a shallow or flat gradient in \logU. However, \cite{rosalesortega2011} noted an increase with radius beyond $0.3R_{25}$ for the entire galaxy, a trend more evident when they split the galaxy into quadrants. Figure~\ref{gradients} shows their fitted gradients for the north and south quadrants, red dashed and dash-dotted lines, respectively, which contain the largest and brightest \hii\ regions in the galaxy and are the most populated by number in their study. The positive gradients seen here could represent any number of changing physical conditions with radius (see \citealt{rosalesortega2011} for a discussion). 

Finally, in the bottom panel of Figure~\ref{gradients}, we show the N/O radial gradient. Notably we recover a similar gradient within the uncertainties to those derived with $T_e$ abundances \citep{rosalesortega2011,berg2013,berg2020}. These studies also demonstrated that a piecewise function best describes the N/O gradient, with the N/O ratio flattening at approximately the $R_{25}$. We did not fit a piecewise function to our data due to the paucity of data points beyond $R_{25}$, but the N/O ratio appears to plateau at a value of \num{-1.37 \pm 0.08}. The appearance of this plateau could be an effect of a combination of the primary and secondary nitrogen production in the outermost \hii\ regions \citep{molla2006}. 

Despite the difficult choice of priors, useful line fluxes, and error-weighting, we have recovered the known gradients in O/H, \logU, and N/O for NGC~628. While we infer a steeper O/H gradient than those determined by $T_e$ measurements \citep{rosalesortega2011,berg2013,berg2020}, our gradient is consistent with other strong-line methods that use the same emission lines that we have \citep{moustakas2010}. The \logU\ and N/O gradients agree more readily with other spectroscopic data, even recovering a possible plateau in N/O at large radii. This remarkable ability for NebulaBayes to estimate the physical properties of \hii\ regions in NGC~628 gives confidence in this method, allowing us to turn to correlations between these properties. 

\section{The \logU-O/H Correlation}\label{sec:corr}

Numerous studies have investigated the potential correlation between the ionization parameter and oxygen abundance. \cite{dopita2006} predicted an anti-correlation of the form $\mathcal{U} \propto Z^{-0.8}$ based on a theoretical calculation of a wind-driven bubble model for \hii\ regions. More recent studies using a large sample of \hii\ regions or star-forming galaxies have since supported this theoretical expectation \citep{maier2006,nagao2006,perezmontero2014,morisset2016,thomas2019}. However, other studies have found there to be either no correlation or a positive correlation \citep[e.g.,][]{dors2011,dopita2014,poetrodjojo2018,kreckel2019,mingozzi2020,ji2022} in direct contrast to \cite{dopita2006}. 

\begin{figure*}
\includegraphics[width=\textwidth,keepaspectratio]{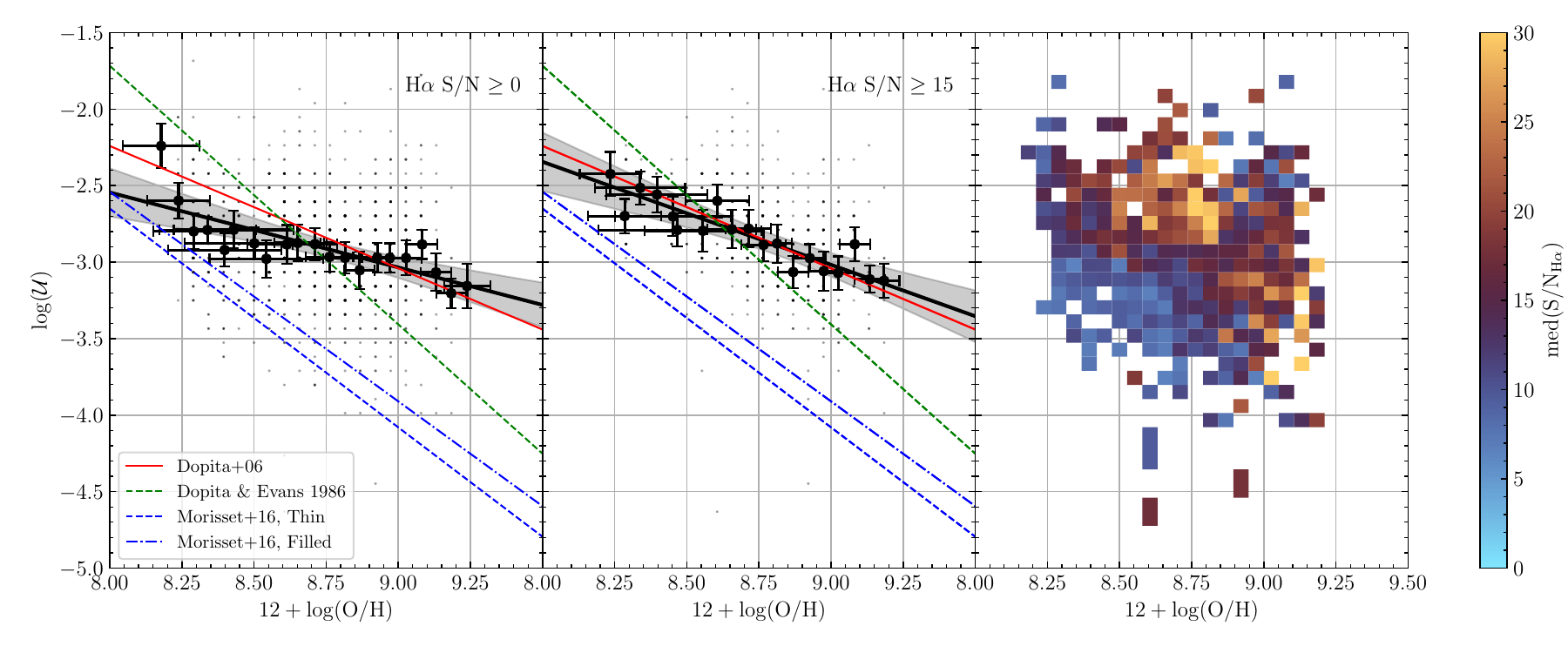}
\caption{The \logU-O/H relation. Left: all \hii\ regions. Middle: only \hii\ regions with \ha\ $\text{S/N} \geq 15$. Individual \hii\ regions are shown as small points which are binned in O/H with a bin width of \SI{0.05}{\dex}. Uncertainties are estimated using a Monte Carlo algorithm. The black line is the fit to the binned data. Relations from the literature are also plotted: \cite{dopita2006} in red, \cite{dopita1986} in dashed green, and the thin shell model (dashed blue) and filled sphere model (dash-dot blue) from \citetalias{morisset2016}. Fits are given in Table~\ref{fits}. Right: 2D histogram of the median \ha\ S/N in each bin.}
\label{MI-corr}
\end{figure*}

Figure~\ref{MI-corr} shows the relation between \logU\ and oxygen abundance for NGC~628 using \ac{signals} data. In the left panel, we show all \hii\ regions in our dataset, while in the middle panel, we show only those with an \ha\ $\text{S/N} \geq 15$ which consists of \SI{47}{\percent} of the total number of regions. The right panel displays a 2D histogram of the median S/N in each bin. Similarly to fitting the gradients in Section~\ref{sec:gradients}, we binned the data in both panels in equally spaced abundance bins of \SI{0.05}{\dex} estimating uncertainties in the bins using a Monte Carlo algorithm run $1000$ times. Given the uncertainties in both the $x$- and $y$-directions, we fit these binned data points using \texttt{scipy.odr}, an implementation of orthogonal distance regression, weighting the fits on each axis by the standard deviations within each bin. In the fit to all data, we find a relation of the form $U \propto Z^{\num{-0.50 \pm 0.09}}$, while for the fit to the high S/N sample gives a relation of $U \propto Z^{\num{-0.67 \pm 0.08}}$. Table~\ref{fits} gives the full parameterization of our fits in both panels.

Additionally, we added the theoretical relations from \cite{dopita1986} and \cite{dopita2006} as well as the relation of \citetalias{morisset2016} derived from photoionization modeling. Since \cite{dopita2006} only reported a proportionality, namely $\mathcal{U} \propto Z^{-0.8}$, in order to plot it, we fit this function to our binned data in each panel while fixing the slope to $-0.8$. This fit is easily done in \texttt{scipy.odr} using the \texttt{ifixb} argument. In both cases, the resulting $y$-intercept is identical within the uncertainties. 

We easily find a negative anti-correlation between \logU\ and O/H in agreement with the predictions of \cite{dopita2006}. Both sample selections show a statistically significant anti-correlation in the binned data (Table~\ref{fits}), with the higher S/N sample matching that of \cite{dopita2006} within the uncertainties. Note that we did not assume these two quantities would be anti-correlated in our \textsc{cloudy} models or NebulaBayes parameter estimation. The fact that we still find this anti-correlation suggests a physical connection between the two parameters.

This finding is in direct tension with those studies that did not find an anti-correlation for individual galaxies \citep{garnett1997,dors2011,dopita2014,poetrodjojo2018,mingozzi2020,grasha2022}. \citetalias{kewley2019} proposes that those studies that do not find an anti-correlation are generally spatially resolved studies. However, we suggest that these are not truly spatially resolved as they could not have resolved individual \hii\ regions at their galaxies' distances. For instance, the best spatial resolution of these studies was that of \cite{dopita2014} with an average spatial resolution of $\sim$\SI{460}{\parsec}. Others had even higher spatial resolutions up to $\sim$\SI{2}{\kilo\parsec} \citep{poetrodjojo2018,mingozzi2020}. While \hii\ regions do come in a wide variety of physical sizes, taking \SI{100}{\parsec} as a reference value \citep{azimlu2011}, means that many of these previous studies could not have resolved individual \hii\ regions and were likely blending multiple \hii\ regions together. 

Compare these resolutions to the \ac{sitelle} (seeing limited) angular resolution of \ang{;;1}, which at NGC~628 corresponds to a spatial resolution of \SI{35}{\parsec}. Our smaller resolution allows us to measure the emission-line properties of small \hii\ regions that simply cannot be measured by other studies while also allowing us to account for all of the emission from larger \hii\ regions. However, \cite{kreckel2019} used data for NGC~628 taken from the MUSE spectrograph which has a similar spatial resolution to \ac{sitelle} and they find a positive correlation between abundance and ionization parameter.\footnote{\cite{kreckel2019} did not report any linear fits to the \logU-O/H correlation they found, only noting that the correlation is positive and reporting the correlation coefficient. See Fig.~6 in \cite{kreckel2019}.} It is worth mentioning how our two studies differ and how that might lead to our two contrary findings. 

First, the spatial coverage of MUSE is limited to only the inner $0.5R_{25}$ of NGC~628. If we limit our data to only the inner $0.5R_{25}$ and plot the resulting \logU-O/H correlation, we recover a weak positive correlation with a slope of \num{0.11 \pm 0.08} with a Spearman $\rho$ coefficient of $+0.16$ at $p = \num{0.51}$. NGC~628 was one of the galaxies in their sample for which this trend was weak; they measured a Spearman coefficient of $+0.19$ between oxygen abundance and \Usulf. 

A few more considerations preclude a direct comparison between our samples. The spectral coverage of MUSE does not include the \oii\ doublet. To estimate oxygen abundance, they used the $S$-calibration of \cite{pilyugin2016}, which utilizes the \nii, \sii, and \oiii\ emission lines and shows minor differences compared to auroral line estimates. The effect of this calibration is a shallow abundance gradient (Figure~\ref{gradients}) with less dynamic range; they report abundances in the range of $\sim$\SIrange{8.4}{8.6}{\dex}. 

\begin{figure*}[ht!]
\includegraphics[width=\textwidth,keepaspectratio]{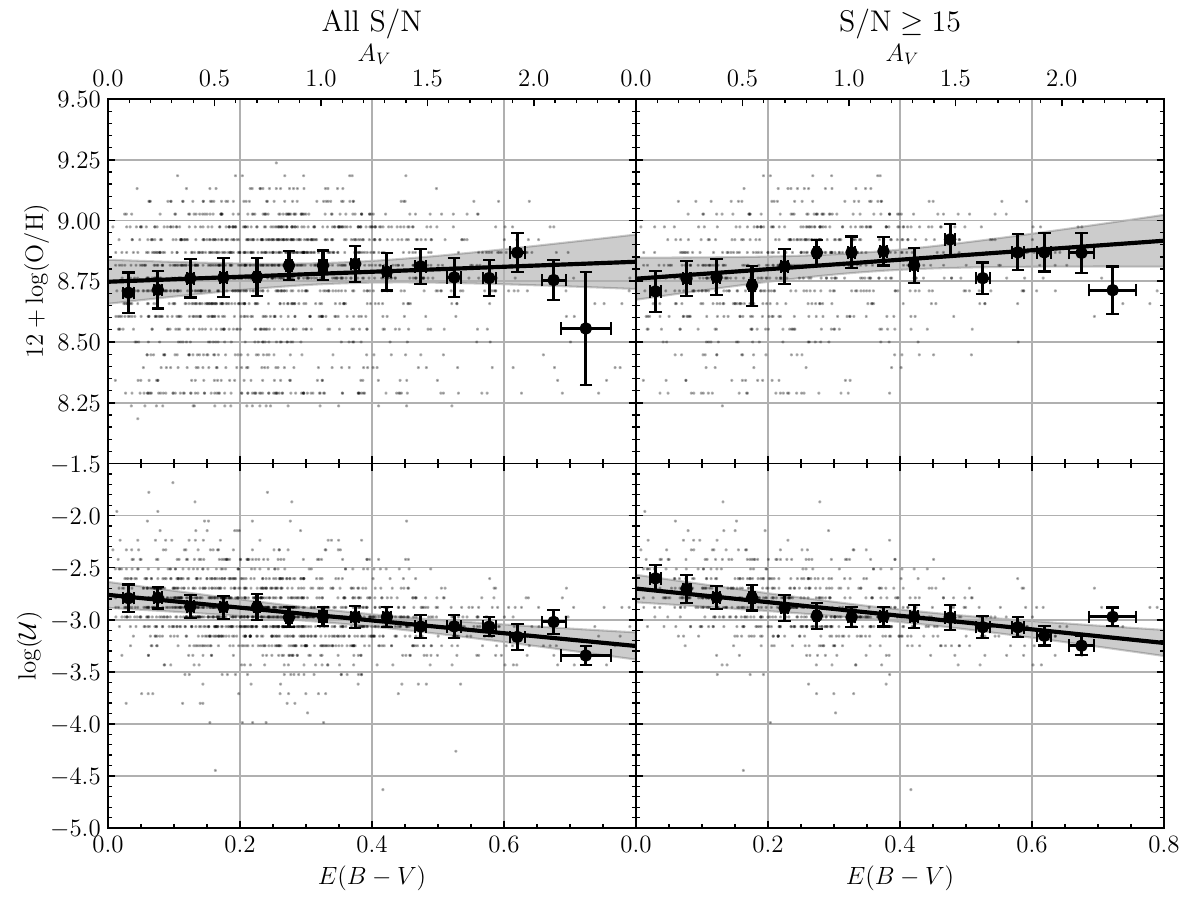}
\caption{The relationship between O/H or \logU\ and dust extinction. The bottom $x$-axis is the color excess, $E(B-V)$, while the top $x$-axis shows extinction, $A_V$, assuming Milky Way reddening. Left column: all \hii\ regions. Right column: only \hii\ regions with \ha\ $\text{S/N} \geq 15$. Individual \hii\ regions are shown as small points which are binned in $E(B-V)$ with a bin width of $0.05$. Uncertainties are estimated using a Monte Carlo algorithm. The black line is a fit to the binned data; parameterization is reported in Table~\ref{fits}.}
\label{ext-grads}
\end{figure*}

Furthermore, \cite{kreckel2019} does avoid the problems with various \U\ calibrations by only comparing oxygen abundance to \Usulf. However, they did not correct any of their emission line measurements for the \ac{dig}, which should elevate the integrated \nii\ and \sii\ fluxes. Not making this correction has several compounding effects. When blended with the emission from an \hii\ region, the \ac{dig} can: artificially flatten metallicity gradients \citep{zhang2017,poetrodjojo2019}, lead to a misclassification of regions in the BPT diagrams \citep{congiu2023}, and lead to an underestimate of \Usulf\ \citep{belfiore2022}. These all likely strongly affect the resulting \logU-O/H correlation that they measure. For instance, \cite{kreckel2019} estimate that a \ac{dig} correction would increase by $\sim$\SI{70}{\percent} their \Usulf\ ratios. Given our extensive spatial coverage, \ac{dig} corrections, and self-consistent estimates of oxygen abundance and ionization parameter, the anti-correlation found in Figure~\ref{MI-corr} is likely robust against comparing our two studies. A forthcoming paper will present a direct comparison of the MUSE \hii\ regions and \ac{signals} \hii\ regions in NGC~628 (J.\ Vandersnickt et al., in prep.).

Despite the anti-correlation between \logU\ and O/H that we find in the binned data, there is still a significant scatter in the raw, unbinned data. In other words, at any fixed oxygen abundance, there is a large spread in \logU\ between \SIrange{1}{2.5}{\dex}. This scatter implies that factors other than oxygen abundance drive the variation in \logU\ within the sample.

\subsection{Internal Dust Extinction in \hii\ Regions}

We cannot ignore the impact of dust on the ionizing spectrum seen by the gas cloud. Dust will change the temperature structure of a cloud since it provides alternate heating and cooling mechanisms. However, an \hii\ region must be cool enough to allow for the survival of dust grains, so any dust effects should only matter in high abundance \hii\ regions where cooling mechanisms dominate and dust can survive \citep{draine2011}. Meanwhile, the wavelength dependence of dust extinction will naturally lead to a softening of the ionizing spectrum. Dust extinction usually explains why \U\ generally does not get higher than approximately $-2$ \citep{dopita2002,yeh2012}. Therefore, we expect two trends: as dust extinction increases, oxygen abundance should increase while \U\ decreases. 

Figure~\ref{ext-grads} shows the trend with dust extinction as measured by the color excess, $E(B-V)$, for both oxygen abundance and \logU\ separated by \ha\ S/N in the left and right columns. In all cases, we binned the data in equally spaced bins of $E(B-V)$ with widths of \SI{0.05}{\dex} and used a Monte Carlo algorithm to estimate the uncertainties in the bins. We again used \texttt{scipy.odr} to fit these binned data points, weighting by the standard deviations in each bin. Table~\ref{fits} gives the full parameterizations. 

We see a shallow positive correlation between \mbox{$E(B-V)$} and O/H with a slope of \num{0.11 \pm 0.05} and a strong negative correlation between $E(B-V)$ and \logU\ with a slope of \num{-0.61 \pm 0.08} when fitting all \hii\ regions. The shallow slope of the O/H-$E(B-V)$ relation is concerning, but this is likely a S/N effect. The right column of Figure~\ref{ext-grads} shows the fit for only \hii\ regions with an \ha\ $\text{S/N} \geq 15$ where we recover a slightly stronger positive correlation between $E(B-V)$ and oxygen abundance with a slope of \num{0.19 \pm 0.07}. Still, NGC~628 does have a relatively constant dust-to-gas ratio \citep{kahre2018,vilchez2019}, implying a relatively flat O/H-$E(B-V)$ relation. Meanwhile, the trend with \logU\ remains strongly negative regardless of the S/N bin with a slope of \num{-0.65 \pm 0.11}, indicating that dust strongly modifies the ionizing spectrum of an \hii\ region. 

\begin{figure}
\includegraphics[width=\columnwidth,keepaspectratio]{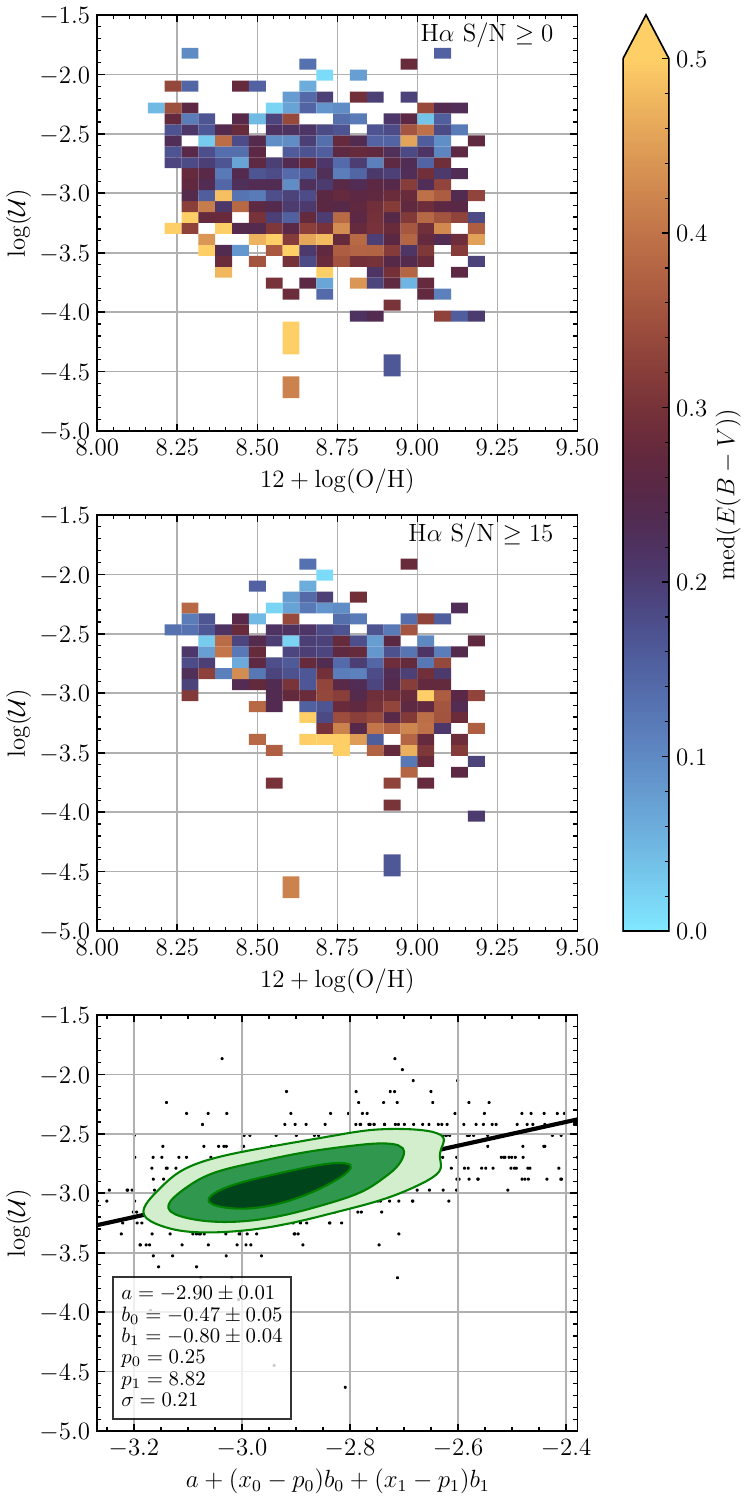}
\caption{Top panel: the 2D histogram of the \logU-O/H anti-correlation with bins colored by dust extinction. This panel includes all points, regardless of H$\alpha$ S/N. Middle panel: same as the top panel except only for those regions with H$\alpha$ $\text{S/N} \geq 15$. Bottom panel: a ``fundamental plane'' combining dust extinction and oxygen abundance on the $x$-axis and \logU\ on the $y$-axis. Here, $x_0 = E(B-V)$ and \mbox{$x_1 = 12 + \log(\mathrm{O/H})$}. The green contours enclose \SIlist{30;50;80}{\percent} of the data points.}
\label{ext-corr}
\end{figure}

As another way of looking at these trends, Figure~\ref{ext-corr}'s top two panels shows a 2D histogram of the \logU-O/H anti-correlation colored by median extinction in each bin. Both panels shows the anti-correlation, for the whole data set (top panel) as well as for high S/N points, $\text{S/N} \geq 15$ (middle panel). We see that, especially for the high S/N data, there is a trend in extinction along the anti-correlation. Namely, those \hii\ regions with high oxygen abundances and low ionization parameters have generally high dust extinction and vice versa. Looking at all data points slightly washes out this trend, but it remains apparent.

Finally, if dust extinction is a significant driver of the anti-correlation between \logU\ and O/H, then we would expect that combining these properties into a ``fundamental plane'' would reduce the scatter compared to the simple two-dimensional anti-correlation. The bottom panel of Figure~\ref{ext-corr} shows such a plane constructed using the \texttt{ltsfit} code \citep{cappellari2013}. This robust method fits a linear function to $n$-dimensional data, accounting for uncertainties in all coordinates and intrinsic scatter. We linearly combine oxygen abundance and extinction against \logU\ for data points with $\text{S/N} \geq 15$. The coefficients for the fit are provided in the panel. We see that the scatter from Figure~\ref{MI-corr} has been reduced by about \SI{0.1}{\dex} by including dust extinction in the correlation. This fit is also statistically significant with a Spearman coefficient of $\rho = +0.54$ at $p = \num{0}$. Thus, dust might be key in understanding the scatter observed in the \logU-O/H anti-correlation. 

An important question to answer is how sensitive our model results are to changes in the Balmer decrement used to derive $E(B-V)$. The Balmer decrement depends on the temperature (and thus the metallicity) and density of an \hii\ region \citep{osterbrock2006}. Using the range of electron temperatures found by \ac{chaos} for NGC~628 and the range of densities we measure, we calculated the Balmer decrement given this grid of temperatures and densities using PyNeb, resulting in a minimum Balmer decrement of \num{2.79} at low abundances and a maximum of \num{3.04} at high abundances. We recalculated the dust extinction, abundances, and ionization parameters, assuming these new Balmer decrement extremes. We found that while the individual properties of any \hii\ region might change, the overall bulk properties of the sample did not. In other words, we found the same trends with extinction and the same anti-correlation as we had before assuming a constant Balmer decrement of $2.86$. Therefore, our conclusions are not sensitive to changes in the Balmer decrement. 

All of this points towards the important consequences dust has on the observed ionizing spectrum of an \hii\ region. For instance, several analytical and numerical models \citep[e.g.,][]{petrosian1972,spitzer1978,arthur2004,haworth2015,ali2021} predict that as the dust opacity of an \hii\ region increases, the size of the \hii\ region should shrink. This is explained by dust absorbing ionizing photons, reducing the pressure gradient between the ionized and neutral gas, and shrinking the \hii\ region compared to one with no dust \citep{ali2021}. Decreasing the size while holding all else constant would raise the ionization parameter according to Equation~\ref{eq:U-strom}. Thus, we would expect the opposite of our \logU-$E(B-V)$ trends here. However, the ionization parameter defined at the Str\"{o}mgren radius as in Equation~\ref{eq:U-strom} would be outside the \hii\ region, so this might require a redefinition of the ionization parameter \citep{ji2022}. The impact of dust on the geometry of an \hii\ region and what it means for the \logU-O/H anti-correlation is beyond the scope of this paper as it requires sophisticated combinations of dynamical and photoionization models of dusty \hii\ regions.

\subsection{\ha\ Surface Brightness}

Star formation plays a vital role in regulating the chemical enrichment of a galaxy. It is well-known that a correlation exists between the global metallicity and stellar mass in star-forming galaxies, i.e., the mass-metallicity relation \citep{lequeux1979,tremonti2004}. There may also be an additional dependence on the \ac{sfr} \citep[e.g.,][]{laralopez2010,mannucci2010}. Spatially resolved studies have shown that this relationship still holds on smaller scales for individual star-forming regions \citep{rosalesortega2012}. 

What is less clear is if the ionization parameter also depends on \ac{sfr}. Some studies have reported a possible correlation between ionization and \ac{sfr}. \cite{dopita2014} found such a correlation for a set of ten star-bursting luminous infrared galaxies, and other studies have found a similar correlation in the low- and high-redshift universe \citep{kaplan2016,reddy2023a,reddy2023b}. One possible way to relate the ionization parameter to the \ac{sfr} is through cluster mass, to which \cite{dopita2006} found \U\ is related. Following the derivation in \cite{ji2022}, if \U\ is regulated through the \ac{sfr} surface density via $\mathcal{U} \propto \Sigma_{\text{SFR}}^\alpha$ with the cluster mass being the medium, then $\alpha < 0.2$. 

\begin{figure*}
\includegraphics[width=\textwidth,keepaspectratio]{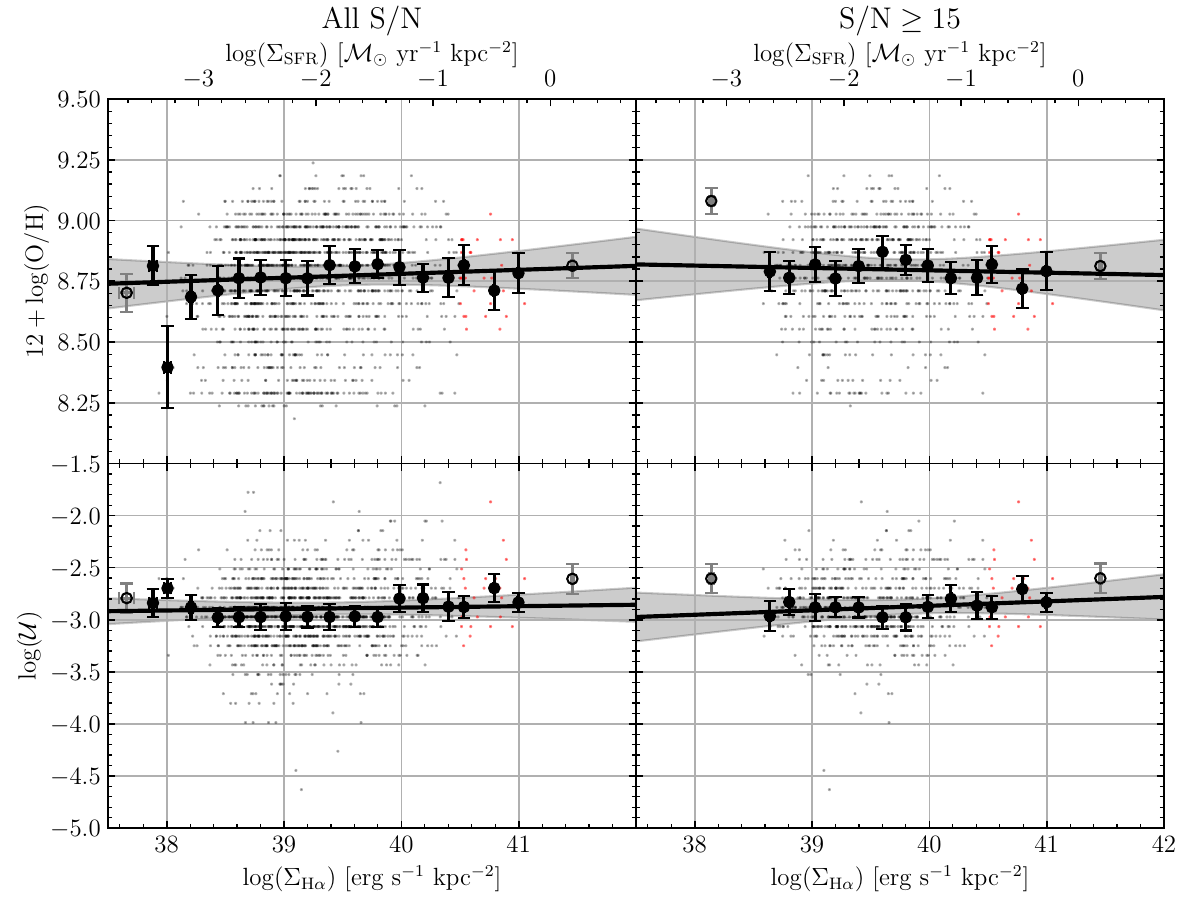}
\caption{The O/H-$\log\Sigma_{\text{\ha}}$ and \logU-$\log\Sigma_{\text{\ha}}$ relations. The bottom $x$-axis is \ha\ surface density while the top $x$-axis is the \ac{sfr} surface density. Left column: all \hii\ regions. Right colum: only \hii\ regions with \ha\ $\text{S/N} \geq 15$. Individual \hii\ regions are shown as small points which are binned in $\log\Sigma_{\text{\ha}}$ with a bin width of \SI{0.2}{dex}. Those regions that satisfy the conditions explore in \cite{dopita2014} are colored red; these are still included in the bins. Large black points with error bars are fitted, while those bins containing only 1 \hii\ region are shown as transparent points and are not included in the fit. Uncertainties are estimated using a Monte Carlo algorithm. The black line is a fit to the solid binned data; the parameterization is reported in Table~\ref{fits}. }
\label{surf-grads}
\end{figure*}

Figure~\ref{surf-grads} shows the spatially resolved $\Sigma_{\text{H}\alpha}$ correlations of O/H and \logU\ for our \hii\ regions separated into two \ha\ S/N samples, all \hii\ regions (left column) and those with \ha\ $\text{S/N} \geq 15$ (right column). We used the extinction-corrected $\log\Sigma_{\text{H}\alpha}$ to represent $\log\Sigma_{\text{SFR}}$ as they differ only be a constant \citep{hao2011,murphy2011,kennicutt2012}, although we show the corresponding \ac{sfr} surface density along the top axis. In all cases, we binned the data in equally spaced bins of \SI{0.2}{\dex} and again used a Monte Carlo algorithm to estimate the bin uncertainties. Bins with only one \hii\ region appear as transparent circles in Figure~\ref{surf-grads}, and the fitting algorithm does not include them. Table~\ref{fits} provides the full parameterizations of the fits.

Regardless of the S/N sample, both relations show very shallow slopes that are consistent with being flat. While this means we do have $\alpha < 0.2$, given the lack of statistical significance, it is unlikely that the very weak O/H-\ac{sfr} and \logU-\ac{sfr} relations could give rise to the stronger \logU-O/H anti-correlation. Not finding a relationship between these two quantities is perhaps not unexpected since NGC~628 is not a starburst galaxy. Comparing our data to that of \cite{dopita2014}, while we do have \hii\ regions with $\log\mathcal{U} \gtrsim -3.3$, we do not have extremely star-forming regions. Using their Figure~13 as a guide, we observe that their positive correlation starts at approximately $\log\Sigma_{\text{SFR}} \simeq -0.5$ or $\log\Sigma_{\text{\ha}} \simeq 40.5$. Only 31 \hii\ regions have such high ionization parameters and luminosity surface densities in our sample (colored red in Figure~\ref{surf-grads}). Interestingly, if we fit only these 31 \hii\ regions without binning, we recover a fit of the form $\mathcal{U} \propto \Sigma_{\text{SFR}}^{0.32 \pm 0.24}$, very similar to \cite{dopita2014} who found an exponent of \num{0.34 \pm 0.08}. This agreement is encouraging, but our lack of data points at these extreme conditions prevent us from making any robust conclusions about the role of luminosity surface density as a mediator of the main anti-correlation.

\section{Conclusions}\label{sec:conclusion}

In this work, we have analyzed the properties over \num{1500} \hii\ regions in the well-known, nearly face-on spiral galaxy NGC~628. The data were acquired using the \ac{cfht} spectro-imageur \ac{sitelle}, which provides a spatial resolution of $\sim$\SI{35}{\parsec} and a spectral resolution ranging from $R \sim 600$ in the blue ($\sim$\SI{3745}{\angstrom}) up to $R \sim 1800$ in the red ($\sim$\SI{6670}{\angstrom}). The instrument's sensitivity enabled us to detect very faint \hii\ regions, down to \SI{38.5}{\erg\per\second\per\square\kilo\parsec}. Its extensive field of view, covering \SI{11}{\square\arcmin}, encompassed the entire optical disk, including contributions from the diffuse ionized gas. This comprehensive coverage allowed us to investigate the potential correlation between the gas-phase oxygen abundance and the ionization parameter. 


Just as is the case when estimating oxygen abundances, several calibrations for the ionization parameter exist in the literature. Unfortunately, as we have shown, these ionization parameter calibrations have the same issues as the oxygen abundance calibrations: differing atomic data sets, stellar \acp{sed}, and chemical abundance sets that results in offsets of \SI{0.5}{\dex} or more. We strongly advise against blindly using these published calibrations without understanding their inherent assumptions and resulting uncertainties. 

These issues motivated us to create our own custom photoionization models using the latest version of \textsc{cloudy} \citep{chatzikos2023}. This allowed us to match the input parameters of the models as best as possible to the observed properties of our \hii\ regions. We used the Bayesian inference code NebulaBayes \citep{thomas2018} to subsequently match those models to our \hii\ regions to derive oxygen abundances, nitrogen abundances, and ionization parameters for each individual \hii\ region. Our ability to reproduce the well-known gradients in NGC~628, consistent with those previously found in the literature \citep[e.g.,][]{rosalesortega2011,berg2020}, gives us confidence in the robustness of our methodology. 

With our estimated physical parameters, we found an anti-correlation between the oxygen abundance and ionization parameter consistent with the theoretical predictions of a wind-driven bubble model for an \hii\ region \citep{dopita2006}. Namely, we recover an anti-correlation of the form $\mathcal{U} \propto Z^{-0.67 \pm 0.08}$ in those regions with high signal-to-noise. We claim that our results, which stand in tension with other studies, are found because we are able to resolve emission from \hii\ regions from a wide variety of physical scales due to our small angular resolution, $\sim$\ang{;;1}. \cite{kreckel2019} used data taken from the MUSE spectrograph, which shares the same spatial resolution as \ac{signals} at NGC~628, and found a positive correlation. While there are some differences in our data---field of view, \hii\ region extraction, and \ac{dig} correction---these variations present an exciting opportunity for further exploration of our findings. Work comparing our two catalogs is ongoing (J.\ Vandersnickt et al., in prep.). 

However, there is still a high degree of scatter in this relationship. Searching for a secondary variable that could  be the cause, we investigated potential trends with the dust extinction as measured through the color excess, $E(B-V)$, and the star formation rate surface density. We found strong trends with dust extinction in both ionization parameter and oxygen abundance: as the oxygen abundance increases and the ionization parameter decreases, the dust extinction increases on average. This is likely due to the well-known connection between dust and metallicity \citep[e.g.,][]{draine2011} where only metal-rich \hii\ regions are cool enough for dust to survive, and that dust absorbs ionizing photons which reduces the ionization parameter. Incorporating dust extinction into a ``fundamental plane'' analysis of the O/H-\logU\ correlation reduces scatter by about \SI{0.1}{\dex} and shows a significant correlation, highlighting dust's importance in understanding the anti-correlation. Meanwhile, we found no trends between either physical property with the star formation rate surface density. This is perhaps not expected, since NGC~628 is not a starburst galaxy in which this trend has been found in the past \citep{dopita2014}. 

As a concluding remark, we note that this research is only a first step. The nature of the correlation between oxygen abundance and ionization parameter is still an open question, one which we hopefully have provided more insight to. As the \ac{signals} team collects and reduces more data, we will add to the sample studied here, improving our ability to make statistical statements about multiple galaxies rather than only one.

\begin{acknowledgments}
The authors would like to thank the referee for helpful comments that improved the readability of the paper. R.G.\ would like to thank Justin Spilker for interesting and helpful discussions. This paper is based on data obtained for \ac{signals}, a large program conducted at the \acf{cfht}, which is operated by the National Research Council of Canada, the Institut National des Sciences de l'Univers of the Centre Naitonal de la Recherche Scientifique of France, and the University of Hawaii. The observations were objected with \ac{sitelle}, a joint project of Universit\'{e} Laval, ABB, Universit\'{e}  de Montr\'{e}al and the \ac{cfht}, with support from the Canada Foundation for Innovation, the National Sciences and Engineering Research Council of Canada (NSERC), and the Fonds de Recherche du Qu\'{e}bec--Nature et Technologies (FRQNT). The authors wish to recognize and acknowledge the very significant cultural role that the summit of Mauna Kea has always had within the indigenous Hawaiian community. We are most grateful to have the opportunity to conduct observations from this mountain. 
\end{acknowledgments}

\facility{CFHT}

\software{\texttt{Astropy} \citep{astropy1,astropy2,astropy3}, \texttt{Matplotlib} \citep{hunter2007}, \texttt{NumPy} \citep{harris2020}, \texttt{SciPy} \citep{virtanen2020}, \texttt{delta\_method}\footnote{\url{https://github.com/gjpelletier/delta_method/tree/main}}, NebulaBayes \citep{thomas2018}, \textsc{cloudy} \citep{chatzikos2023}, \texttt{ltsfit} \citep{cappellari2013}, \texttt{cmcrameri} \citep{crameri2018}}

\bibliography{SIGNALS_OH-logU.bib}

\begin{thebibliography}{}
\expandafter\ifx\csname natexlab\endcsname\relax\def\natexlab#1{#1}\fi
\providecommand{\url}[1]{\href{#1}{#1}}
\providecommand{\dodoi}[1]{doi:~\href{http://doi.org/#1}{\nolinkurl{#1}}}
\providecommand{\doeprint}[1]{\href{http://ascl.net/#1}{\nolinkurl{http://ascl.net/#1}}}
\providecommand{\doarXiv}[1]{\href{https://arxiv.org/abs/#1}{\nolinkurl{https://arxiv.org/abs/#1}}}

\bibitem[{Ali(2021)}]{ali2021}
Ali, A.~A. 2021, MNRAS, 501, 4136, \dodoi{10.1093/mnras/staa3992}

\bibitem[{Ali {et~al.}(1991)Ali, Blum, Bumgardner, Cranmer, Ferland, Haefner,
  \& Tiede}]{ali1991}
Ali, B., Blum, R.~D., Bumgardner, T.~E., {et~al.} 1991, PASP, 103, 1182,
  \dodoi{10.1086/132938}

\bibitem[{{Allende Prieto} {et~al.}(2001){Allende Prieto}, Lambert, \&
  Asplund}]{allendeprieto2001}
{Allende Prieto}, C., Lambert, D.~L., \& Asplund, M. 2001, ApJL, 556, L63,
  \dodoi{10.1086/322874}

\bibitem[{Aller(1942)}]{aller1942}
Aller, L.~H. 1942, ApJ, 95, 52, \dodoi{10.1086/144372}

\bibitem[{Arthur {et~al.}(2004)Arthur, Kurtz, Franco, \&
  Albarr\'{a}n}]{arthur2004}
Arthur, S.~J., Kurtz, S.~E., Franco, J., \& Albarr\'{a}n, M.~Y. 2004, ApJ, 608,
  282, \dodoi{10.1086/386366}

\bibitem[{Asplund {et~al.}(2009)Asplund, Grevesse, Sauval, \&
  Scott}]{asplund2009}
Asplund, M., Grevesse, N., Sauval, A.~J., \& Scott, P. 2009, ARA\&A, 47, 481,
  \dodoi{10.1146/annurev.astro.46.060407.145222}

\bibitem[{{Astropy Collaboration} {et~al.}(2013){Astropy Collaboration},
  Robitaille, Tollerud, Greenfield, Droettboom, Bray, Aldcroft, Davis,
  Ginsburg, Price-Whelan, Kerzendorf, Conley, Crighton, Barbary, Muna,
  Ferguson, Grollier, Parikh, Nair, Unther, Deil, Woillez, Conseil, Kramer,
  Turner, Singer, Fox, Weaver, Zabalza, Edwards, {Azalee Bostroem}, Burke,
  Casey, Crawford, Dencheva, Ely, Jenness, Labrie, Lim, Pierfederici, Pontzen,
  Ptak, Refsdal, Servillat, \& Streicher}]{astropy1}
{Astropy Collaboration}, Robitaille, T.~P., Tollerud, E.~J., {et~al.} 2013,
  A\&A, 558, A33, \dodoi{10.1051/0004-6361/201322068}

\bibitem[{{Astropy Collaboration} {et~al.}(2018){Astropy Collaboration},
  Price-Whelan, Sipőcz, Günther, Lim, Crawford, Conseil, Shupe, Craig,
  Dencheva, Ginsburg, VanderPlas, Bradley, Pérez-Suárez, de~Val-Borro,
  Aldcroft, Cruz, Robitaille, Tollerud, Ardelean, Babej, Bach, Bachetti,
  Bakanov, Bamford, Barentsen, Barmby, Baumbach, Berry, Biscani, Boquien,
  Bostroem, Bouma, Brammer, Bray, Breytenbach, Buddelmeijer, Burke, Calderone,
  Rodríguez, Cara, Cardoso, Cheedella, Copin, Corrales, Crichton, D’Avella,
  Deil, Depagne, Dietrich, Donath, Droettboom, Earl, Erben, Fabbro, Ferreira,
  Finethy, Fox, Garrison, Gibbons, Goldstein, Gommers, Greco, Greenfield,
  Groener, Grollier, Hagen, Hirst, Homeier, Horton, Hosseinzadeh, Hu, Hunkeler,
  Ivezić, Jain, Jenness, Kanarek, Kendrew, Kern, Kerzendorf, Khvalko, King,
  Kirkby, Kulkarni, Kumar, Lee, Lenz, Littlefair, Ma, Macleod, Mastropietro,
  McCully, Montagnac, Morris, Mueller, Mumford, Muna, Murphy, Nelson, Nguyen,
  Ninan, Nöthe, Ogaz, Oh, Parejko, Parley, Pascual, Patil, Patil, Plunkett,
  Prochaska, Rastogi, Janga, Sabater, Sakurikar, Seifert, Sherbert,
  Sherwood-Taylor, Shih, Sick, Silbiger, Singanamalla, Singer, Sladen, Sooley,
  Sornarajah, Streicher, Teuben, Thomas, Tremblay, Turner, Terrón, Kerkwijk,
  de~la Vega, Watkins, Weaver, Whitmore, Woillez, \& Zabalza}]{astropy2}
{Astropy Collaboration}, Price-Whelan, A.~M., Sipőcz, B.~M., {et~al.} 2018,
  AJ, 156, 123, \dodoi{10.3847/1538-3881/aabc4f}

\bibitem[{{Astropy Collaboration} {et~al.}(2022){Astropy Collaboration},
  Price-Whelan, Lim, Earl, Starkman, Bradley, Shupe, Patil, Corrales, Brasseur,
  Nöthe, Donath, Tollerud, Morris, Ginsburg, Vaher, Weaver, Tocknell,
  Jamieson, van Kerkwijk, Robitaille, Merry, Bachetti, Günther, Aldcroft,
  Alvarado-Montes, Archibald, Bódi, Bapat, Barentsen, Bazán, Biswas, Boquien,
  Burke, Cara, Cara, Conroy, Conseil, Craig, Cross, Cruz, D’Eugenio,
  Dencheva, Devillepoix, Dietrich, Eigenbrot, Erben, Ferreira, Foreman-Mackey,
  Fox, Freij, Garg, Geda, Glattly, Gondhalekar, Gordon, Grant, Greenfield,
  Groener, Guest, Gurovich, Handberg, Hart, Hatfield-Dodds, Homeier,
  Hosseinzadeh, Jenness, Jones, Joseph, Kalmbach, Karamehmetoglu,
  Kałuszyński, Kelley, Kern, Kerzendorf, Koch, Kulumani, Lee, Ly, Ma,
  MacBride, Maljaars, Muna, Murphy, Norman, O’Steen, Oman, Pacifici, Pascual,
  Pascual-Granado, Patil, Perren, Pickering, Rastogi, Roulston, Ryan, Rykoff,
  Sabater, Sakurikar, Salgado, Sanghi, Saunders, Savchenko, Schwardt,
  Seifert-Eckert, Shih, Jain, Shukla, Sick, Simpson, Singanamalla, Singer,
  Singhal, Sinha, Sipőcz, Spitler, Stansby, Streicher, Šumak, Swinbank,
  Taranu, Tewary, Tremblay, Val-Borro, Van~Kooten, Vasović, Verma,
  de~Miranda~Cardoso, Williams, Wilson, Winkel, Wood-Vasey, Xue, Yoachim,
  Zhang, \& Zonca}]{astropy3}
{Astropy Collaboration}, Price-Whelan, A.~M., Lim, P.~L., {et~al.} 2022, ApJ,
  935, 167, \dodoi{10.3847/1538-4357/ac7c74}

\bibitem[{Azimlu {et~al.}(2011)Azimlu, Marciniak, \& Barmby}]{azimlu2011}
Azimlu, M., Marciniak, R., \& Barmby, P. 2011, AJ, 142, 139,
  \dodoi{10.1088/0004-6256/142/4/139}

\bibitem[{Badnell {et~al.}(2015)Badnell, Ferland, Gorczyca, Nikoli\'{c}, \&
  Wagle}]{badnell2015}
Badnell, N.~R., Ferland, G.~J., Gorczyca, T.~W., Nikoli\'{c}, D., \& Wagle,
  G.~A. 2015, ApJ, 804, 100, \dodoi{10.1088/0004-637x/804/2/100}

\bibitem[{Baldwin {et~al.}(1981)Baldwin, Phillips, \& Terlevich}]{baldwin1981}
Baldwin, J.~A., Phillips, M.~M., \& Terlevich, R. 1981, PASP, 93, 5,
  \dodoi{10.1086/130766}

\bibitem[{Belfiore {et~al.}(2022)Belfiore, Santoro, Groves, Schinnerer,
  Kreckel, Glover, Klessen, Emsellem, Blanc, Congiu, Barnes, Boquien, Chevance,
  Dale, Kruijssen, Leroy, Pan, Pessa, Schruba, \& Williams}]{belfiore2022}
Belfiore, F., Santoro, F., Groves, B., {et~al.} 2022, A\&A, 659, A26,
  \dodoi{10.1051/0004-6361/202141859}

\bibitem[{Berg {et~al.}(2020)Berg, Pogge, Skillman, Croxall, Moustakas, Rogers,
  \& Sun}]{berg2020}
Berg, D.~A., Pogge, R.~W., Skillman, E.~D., {et~al.} 2020, ApJ, 893, 96,
  \dodoi{10.3847/1538-4357/ab7eab}

\bibitem[{Berg {et~al.}(2015)Berg, Skillman, Croxall, Pogge, Moustakas, \&
  Johnson-Groh}]{berg2015}
Berg, D.~A., Skillman, E.~D., Croxall, K.~V., {et~al.} 2015, ApJ, 806, 16,
  \dodoi{10.1088/0004-637X/806/1/16}

\bibitem[{Berg {et~al.}(2013)Berg, Skillman, Garnett, Croxall, Marble, Smith,
  Gordon, \& Kennicutt}]{berg2013}
Berg, D.~A., Skillman, E.~D., Garnett, D.~R., {et~al.} 2013, ApJ, 775, 128,
  \dodoi{10.1088/0004-637x/775/2/128}

\bibitem[{Blanc {et~al.}(2015)Blanc, Kewley, Vogt, \& Dopita}]{blanc2015}
Blanc, G.~A., Kewley, L.~J., Vogt, F. P.~A., \& Dopita, M.~A. 2015, ApJ, 798,
  99, \dodoi{10.1088/0004-637X/798/2/99}

\bibitem[{Borsenberger \& Stasi\'{n}ska(1982)}]{borsenberger1982}
Borsenberger, J., \& Stasi\'{n}ska, G. 1982, A\&A, 106, 158

\bibitem[{Bresolin {et~al.}(1999)Bresolin, Kennicutt, \&
  Garnett}]{bresolin1999}
Bresolin, F., Kennicutt, R.~C., \& Garnett, D.~R. 1999, ApJ, 510, 104,
  \dodoi{10.1086/306576}

\bibitem[{Cantiello {et~al.}(2007)Cantiello, Yoon, Langer, \&
  Livio}]{cantiello2007}
Cantiello, M., Yoon, S.-C., Langer, N., \& Livio, M. 2007, A\&A, 465, L29,
  \dodoi{10.1051/0004-6361:20077115}

\bibitem[{Cappellari {et~al.}(2013)Cappellari, Scott, Alatalo, Blitz, Bois,
  Bournaud, Bureau, Crocker, Davies, Davis, de~Zeeuw, Duc, Emsellem, Khochfar,
  Krajnovi\'{c}, Kuntschner, McDermid, Morganti, Naab, Oosterloo, Sarzi, Serra,
  Weijmans, \& Young}]{cappellari2013}
Cappellari, M., Scott, N., Alatalo, K., {et~al.} 2013, MNRAS, 432, 1709,
  \dodoi{10.1093/mnras/stt562}

\bibitem[{Cardelli {et~al.}(1989)Cardelli, Clayton, \& Mathis}]{cardelli1989}
Cardelli, J.~A., Clayton, G.~C., \& Mathis, J.~S. 1989, ApJ, 345, 245,
  \dodoi{10.1086/167900}

\bibitem[{Cedr\'{e}s {et~al.}(2013)Cedr\'{e}s, Beckman, Bongiovanni, Cepa,
  {Asensio Ramos}, Giammanco, Cabrera-Lavers, \& Alfaro}]{cedres2013}
Cedr\'{e}s, B., Beckman, J.~E., Bongiovanni, A., {et~al.} 2013, ApJL, 765, L24,
  \dodoi{10.1088/2041-8205/765/1/l24}

\bibitem[{Chatzikos {et~al.}(2023)Chatzikos, Bianchi, Camilloni, Chakraborty,
  Gunasekera, Guzm\'{a}n, Milby, Sarkar, Shaw, van Hoof, \&
  Ferland}]{chatzikos2023}
Chatzikos, M., Bianchi, S., Camilloni, F., {et~al.} 2023, RMxAA, 59, 327,
  \dodoi{10.22201/ia.01851101p.2023.59.02.12}

\bibitem[{Congiu {et~al.}(2023)Congiu, Blanc, Belfiore, Santoro, Scheuermann,
  Kreckel, Emsellem, Groves, Pan, Bigiel, Dale, Glover, Grasha, Egorov, Leroy,
  Schinnerer, Watkins, \& Williams}]{congiu2023}
Congiu, E., Blanc, G.~A., Belfiore, F., {et~al.} 2023, A\&A, 672, A148,
  \dodoi{10.1051/0004-6361/202245153}

\bibitem[{Cowie \& Songaila(1986)}]{cowie1986}
Cowie, L.~L., \& Songaila, A. 1986, ARA\&A, 24, 499,
  \dodoi{10.1146/annurev.aa.24.090186.002435}

\bibitem[{Crameri(2018)}]{crameri2018}
Crameri, F. 2018, Zenodo, \dodoi{10.5281/zenodo.2649252}

\bibitem[{Croxall {et~al.}(2015)Croxall, Pogge, Berg, Skillman, \&
  Moustakas}]{croxall2015}
Croxall, K.~V., Pogge, R.~W., Berg, D.~A., Skillman, E.~D., \& Moustakas, J.
  2015, ApJ, 808, 42, \dodoi{10.1088/0004-637x/808/1/42}

\bibitem[{Croxall {et~al.}(2016)Croxall, Pogge, Berg, Skillman, \&
  Moustakas}]{croxall2016}
---. 2016, ApJ, 830, 4, \dodoi{10.3847/0004-637x/830/1/4}

\bibitem[{Dhungana {et~al.}(2016)Dhungana, Kehoe, Vinko, Silverman, Wheeler,
  Zheng, Marion, Fox, Akerlof, Biro, Borkovits, Cenko, Clubb, Filippenko,
  Ferrante, Gibson, Graham, Hegedus, Kelly, Kelemen, Lee, Marschalko,
  Moln\'{a}r, Nagy, Ordasi, Pal, Sarneczky, Shivvers, Szakats, Szalai,
  Szegedi-Elek, Sz\'{e}kely, Szing, Tak\'{a}ts, \& Vida}]{dhungana2016}
Dhungana, G., Kehoe, R., Vinko, J., {et~al.} 2016, ApJ, 822, 6,
  \dodoi{10.3847/0004-637x/822/1/6}

\bibitem[{D\'{\i}az {et~al.}(2000)D\'{\i}az, Castellanos, Terlevich, \& {Luisa
  Garc\'{\i}a-Vargas}}]{diaz2000}
D\'{\i}az, A.~I., Castellanos, M., Terlevich, E., \& {Luisa
  Garc\'{\i}a-Vargas}, M. 2000, MNRAS, 318, 462,
  \dodoi{10.1046/j.1365-8711.2000.03737.x}

\bibitem[{D\'{\i}az {et~al.}(1991)D\'{\i}az, Terlevich, V\'{\i}lchez, Pagel, \&
  Edmunds}]{diaz1991}
D\'{\i}az, A.~I., Terlevich, E., V\'{\i}lchez, J.~M., Pagel, B. E.~J., \&
  Edmunds, M.~G. 1991, MNRAS, 253, 245, \dodoi{10.1093/mnras/253.2.245}

\bibitem[{D\'{\i}az \& Zamora(2022)}]{diaz2022}
D\'{\i}az, A.~I., \& Zamora, S. 2022, MNRAS, 511, 4377,
  \dodoi{10.1093/mnras/stac387}

\bibitem[{Dinerstein \& Shields(1986)}]{dinerstein1986}
Dinerstein, H.~L., \& Shields, G.~A. 1986, ApJ, 311, 45, \dodoi{10.1086/164753}

\bibitem[{Dopita \& Evans(1986)}]{dopita1986}
Dopita, M.~A., \& Evans, I.~N. 1986, ApJ, 307, 431, \dodoi{10.1086/164432}

\bibitem[{Dopita {et~al.}(2002)Dopita, Groves, Sutherland, Binette, \&
  Cecil}]{dopita2002}
Dopita, M.~A., Groves, B.~A., Sutherland, R.~S., Binette, L., \& Cecil, G.
  2002, ApJ, 572, 753, \dodoi{10.1086/340429}

\bibitem[{Dopita {et~al.}(2014)Dopita, Rich, Vogt, Kewley, Ho, Basurah, Ali, \&
  Amer}]{dopita2014}
Dopita, M.~A., Rich, J., Vogt, F. P.~A., {et~al.} 2014, Ap\&SS, 350, 741,
  \dodoi{10.1007/s10509-013-1753-2}

\bibitem[{Dopita {et~al.}(2013)Dopita, Sutherland, Nicholls, Kewley, \&
  Vogt}]{dopita2013}
Dopita, M.~A., Sutherland, R.~S., Nicholls, D.~C., Kewley, L.~J., \& Vogt, F.
  P.~A. 2013, ApJS, 208, 10, \dodoi{10.1088/0067-0049/208/1/10}

\bibitem[{Dopita {et~al.}(2006)Dopita, Fischera, Sutherland, Kewley, Tuffs,
  Popescu, van Breugel, Groves, \& Leitherer}]{dopita2006}
Dopita, M.~A., Fischera, J., Sutherland, R.~S., {et~al.} 2006, ApJ, 647, 244,
  \dodoi{10.1086/505418}

\bibitem[{Dors {et~al.}(2011)Dors, Krabbe, H\"{a}gele, \&
  P\'{e}rez-Montero}]{dors2011}
Dors, O.~L., Krabbe, A., H\"{a}gele, G.~F., \& P\'{e}rez-Montero, E. 2011,
  MNRAS, 415, 3616, \dodoi{10.1111/j.1365-2966.2011.18978.x}

\bibitem[{Dors {et~al.}(2016)Dors, P\'{e}rez-Montero, H\"{a}gele, Cardaci, \&
  Krabbe}]{dors2016}
Dors, O.~L., P\'{e}rez-Montero, E., H\"{a}gele, G.~F., Cardaci, M.~V., \&
  Krabbe, A.~C. 2016, MNRAS, 456, 4407, \dodoi{10.1093/mnras/stv2995}

\bibitem[{Draine(2011)}]{draine2011}
Draine, B.~T. 2011, Physics of the Interstellar and Intergalactic Medium
  (Princeton University Press), \dodoi{10.1515/9781400839087}

\bibitem[{Drissen {et~al.}(2019)Drissen, Martin, Rousseau-Nepton, Robert,
  Martin, Baril, Prunet, Joncas, Thibault, Brousseau, Mandar, Grandmont, Yee,
  \& Simard}]{drissen2019}
Drissen, L., Martin, T., Rousseau-Nepton, L., {et~al.} 2019, MNRAS, 485, 3930,
  \dodoi{10.1093/mnras/stz627}

\bibitem[{Eldridge {et~al.}(2011)Eldridge, Langer, \& Tout}]{eldridge2011}
Eldridge, J.~J., Langer, N., \& Tout, C.~A. 2011, MNRAS, 414, 3501,
  \dodoi{10.1111/j.1365-2966.2011.18650.x}

\bibitem[{Espinosa-Ponce {et~al.}(2022)Espinosa-Ponce, S\'{a}nchez, Morisset,
  Barrera-Ballesteros, Galbany, Garc\'{\i}a-Benito, Lacerda, \&
  Mast}]{espinosaponce2022}
Espinosa-Ponce, C., S\'{a}nchez, S.~F., Morisset, C., {et~al.} 2022, MNRAS,
  512, 3436, \dodoi{10.1093/mnras/stac456}

\bibitem[{Evans(1991)}]{evans1991}
Evans, I.~N. 1991, ApJS, 76, 985, \dodoi{10.1086/191586}

\bibitem[{Ferguson {et~al.}(1998)Ferguson, Gallagher, \& Wyse}]{ferguson1998}
Ferguson, A. M.~N., Gallagher, J.~S., \& Wyse, R. F.~G. 1998, AJ, 116, 673,
  \dodoi{10.1086/300456}

\bibitem[{Ferland(1989)}]{ferland1989}
Ferland, G.~J. 1989, Hazy, A Brief Introduction to Cloudy 74

\bibitem[{Ferland {et~al.}(1998)Ferland, Korista, Verner, Ferguson, Kingdon, \&
  Verner}]{ferland1998}
Ferland, G.~J., Korista, K.~T., Verner, D.~A., {et~al.} 1998, PASP, 110, 761,
  \dodoi{10.1086/316190}

\bibitem[{Ferland {et~al.}(2013)Ferland, Porter, van Hoof, Williams, Abel,
  Lykins, Shaw, Henney, \& Stancil}]{ferland2013}
Ferland, G.~J., Porter, R.~L., van Hoof, P. A.~M., {et~al.} 2013, RMxAA, 49,
  137, \dodoi{10.3847/2515-5172/ad0e75}

\bibitem[{Garnett(1989)}]{garnett1989}
Garnett, D.~R. 1989, ApJ, 345, 282, \dodoi{10.1086/167904}

\bibitem[{Garnett {et~al.}(1995)Garnett, Dufour, Peimbert, Torres-Peimbert,
  Shields, Skillman, Terlevich, \& Terlevich}]{garnett1995}
Garnett, D.~R., Dufour, R.~J., Peimbert, M., {et~al.} 1995, ApJL, 449, L77,
  \dodoi{10.1086/309620}

\bibitem[{Garnett {et~al.}(1997)Garnett, Shields, Skillman, Sagan, \&
  Dufour}]{garnett1997}
Garnett, D.~R., Shields, G.~A., Skillman, E.~D., Sagan, S.~P., \& Dufour, R.~J.
  1997, ApJ, 489, 63, \dodoi{10.1086/304775}

\bibitem[{Goswami {et~al.}(2024)Goswami, Vilchez, P\'{e}rez-D\'{\i}az, Silva,
  Bressan, \& P\'{e}rez-Montero}]{goswami2024}
Goswami, S., Vilchez, J.~M., P\'{e}rez-D\'{\i}az, B., {et~al.} 2024, A\&A, 685,
  A81, \dodoi{10.1051/0004-6361/202348231}

\bibitem[{Grasha {et~al.}(2022)Grasha, Chen, Battisti, Acharyya, Ridolfo,
  Poehler, Mably, Verma, Hayward, Kharbanda, Poetrodjojo, Seibert, Rich,
  Madore, \& Kewley}]{grasha2022}
Grasha, K., Chen, Q.~H., Battisti, A.~J., {et~al.} 2022, ApJ, 929, 118,
  \dodoi{10.3847/1538-4357/ac5ab2}

\bibitem[{Grevesse {et~al.}(2010)Grevesse, Asplund, Sauval, \&
  Scott}]{grevesse2010}
Grevesse, N., Asplund, M., Sauval, A.~J., \& Scott, P. 2010, Ap\&SS, 328, 179,
  \dodoi{10.1007/978-90-481-9198-7_31}

\bibitem[{Hao {et~al.}(2011)Hao, Kennicutt, Johnson, Calzetti, Dale, \&
  Moustakas}]{hao2011}
Hao, C.-N., Kennicutt, R.~C., Johnson, B.~D., {et~al.} 2011, ApJ, 741, 124,
  \dodoi{10.1088/0004-637x/741/2/124}

\bibitem[{Harris {et~al.}(2020)Harris, Millman, van~der Walt, Gommers,
  Virtanen, Cournapeau, Wieser, Taylor, Berg, Smith, Kern, Picus, Hoyer, van
  Kerkwijk, Brett, Haldane, del R\'{\i}o, Wiebe, Peterson, G\'{e}rard-Marchant,
  Sheppard, Reddy, Weckesser, Abbasi, Gohlke, \& Oliphant}]{harris2020}
Harris, C.~R., Millman, K.~J., van~der Walt, S.~J., {et~al.} 2020, Nature, 585,
  357, \dodoi{10.1038/s41586-020-2649-2}

\bibitem[{Haworth {et~al.}(2015)Haworth, Harries, Acreman, \&
  Bisbas}]{haworth2015}
Haworth, T.~J., Harries, T.~J., Acreman, D.~M., \& Bisbas, T.~G. 2015, MNRAS,
  453, 2277, \dodoi{10.1093/mnras/stv1814}

\bibitem[{Hillier \& Miller(1998)}]{hillier1998}
Hillier, D.~J., \& Miller, D.~L. 1998, ApJ, 496, 407, \dodoi{10.1086/305350}

\bibitem[{Hunter(2007)}]{hunter2007}
Hunter, J.~D. 2007, CSE, 9, 90, \dodoi{10.1109/mcse.2007.55}

\bibitem[{Izotov {et~al.}(2006)Izotov, Stasi\'{n}ska, Meynet, Guseva, \&
  Thuan}]{izotov2006}
Izotov, Y.~I., Stasi\'{n}ska, G., Meynet, G., Guseva, N.~G., \& Thuan, T.~X.
  2006, A\&A, 448, 955, \dodoi{10.1051/0004-6361:20053763}

\bibitem[{Izotov {et~al.}(2009)Izotov, Thuan, \& Wilson}]{izotov2009}
Izotov, Y.~I., Thuan, T.~X., \& Wilson, J.~C. 2009, ApJ, 703, 1984,
  \dodoi{10.1088/0004-637x/703/2/1984}

\bibitem[{Jenkins(1987)}]{jenkins1987}
Jenkins, E.~B. 1987, in Astrophysics and Space Science Library, Vol. 134,
  Interstellar Processes, ed. D.~J. Hollenbach \& H.~A. Thronson (Springer),
  533, \dodoi{10.1007/978-94-009-3861-8_20}

\bibitem[{Jenkins(2009)}]{jenkins2009}
Jenkins, E.~B. 2009, ApJ, 700, 1299, \dodoi{10.1088/0004-637x/700/2/1299}

\bibitem[{Ji \& Yan(2022)}]{ji2022}
Ji, X., \& Yan, R. 2022, A\&A, 659, A112, \dodoi{10.1051/0004-6361/202142312}

\bibitem[{Kaasinen {et~al.}(2018)Kaasinen, Kewley, Bian, Groves, Kashino,
  Silverman, \& Kartaltepe}]{kaasinen2018}
Kaasinen, M., Kewley, L., Bian, F., {et~al.} 2018, MNRAS, 477, 5568,
  \dodoi{10.1093/mnras/sty1012}

\bibitem[{Kahre {et~al.}(2018)Kahre, Walterbos, Kim, Thilker, Calzetti, Lee,
  Sabbi, Ubeda, Aloisi, Cignoni, Cook, Dale, Elmegreen, Elmegreen, Fumagalli,
  Gallagher, Gouliermis, Grasha, Grebel, Hunter, Sacchi, Smith, Tosi, Adamo,
  Andrews, Ashworth, Bright, Brown, Chandar, Christian, de~Mink, Dobbs, Evans,
  Herrero, Johnson, Kennicutt, Krumholz, Messa, Nair, Nota, Pellerin, Ryon,
  Schaerer, Shabani, {Van Dyk}, Whitmore, \& Wofford}]{kahre2018}
Kahre, L., Walterbos, R.~A., Kim, H., {et~al.} 2018, ApJ, 855, 133,
  \dodoi{10.3847/1538-4357/aab101}

\bibitem[{Kaplan {et~al.}(2016)Kaplan, Jogee, Kewley, Blanc, Weinzirl, Song,
  Drory, Luo, \& van~den Bosch}]{kaplan2016}
Kaplan, K.~F., Jogee, S., Kewley, L., {et~al.} 2016, MNRAS, 462, 1642,
  \dodoi{10.1093/mnras/stw1422}

\bibitem[{Kauffmann {et~al.}(2003)Kauffmann, Heckman, Tremonti, Brinchmann,
  Charlot, White, Ridgway, Brinkmann, Fukugita, Hall, \v{Z}. Ivezi\'{c},
  Richards, \& Schneider}]{kauffmann2003}
Kauffmann, G., Heckman, T.~M., Tremonti, C., {et~al.} 2003, MNRAS, 346, 1055,
  \dodoi{10.1111/j.1365-2966.2003.07154.x}

\bibitem[{Kennicutt(1984)}]{kennicutt1984}
Kennicutt, R.~C. 1984, ApJ, 287, 116, \dodoi{10.1086/162669}

\bibitem[{Kennicutt \& Evans(2012)}]{kennicutt2012}
Kennicutt, R.~C., \& Evans, N.~J. 2012, ARA\&A, 50, 531,
  \dodoi{10.1146/annurev-astro-081811-125610}

\bibitem[{Kennicutt \& Garnett(1996)}]{kennicutt1996}
Kennicutt, R.~C., \& Garnett, D.~R. 1996, ApJ, 456, 504, \dodoi{10.1086/176675}

\bibitem[{Kewley \& Dopita(2002)}]{kewley2002}
Kewley, L.~J., \& Dopita, M.~A. 2002, ApJS, 142, 35, \dodoi{10.1086/341326}

\bibitem[{Kewley {et~al.}(2001)Kewley, Dopita, Sutherland, Heisler, \&
  Trevena}]{kewley2001}
Kewley, L.~J., Dopita, M.~A., Sutherland, R.~S., Heisler, C.~A., \& Trevena, J.
  2001, ApJ, 556, 121, \dodoi{10.1086/321545}

\bibitem[{Kewley \& Ellison(2008)}]{kewley2008}
Kewley, L.~J., \& Ellison, S.~L. 2008, ApJ, 681, 1183, \dodoi{10.1086/587500}

\bibitem[{Kewley {et~al.}(2006)Kewley, Groves, Kauffmann, \&
  Heckman}]{kewley2006}
Kewley, L.~J., Groves, B., Kauffmann, G., \& Heckman, T. 2006, MNRAS, 372, 961,
  \dodoi{10.1111/j.1365-2966.2006.10859.x}

\bibitem[{Kewley {et~al.}(2019)Kewley, Nicholls, \& Sutherland}]{kewley2019}
Kewley, L.~J., Nicholls, D.~C., \& Sutherland, R.~S. 2019, ARA\&A, 57, 511,
  \dodoi{10.1146/annurev-astro-081817-051832}

\bibitem[{Kobulnicky \& Kewley(2004)}]{kobulnicky2004}
Kobulnicky, H.~A., \& Kewley, L.~J. 2004, ApJ, 617, 240, \dodoi{10.1086/425299}

\bibitem[{Kreckel {et~al.}(2019)Kreckel, Ho, Blanc, Groves, Santoro,
  Schinnerer, Bigiel, Chevance, Congiu, Emsellem, Faesi, Glover, Grasha,
  Kruijssen, Lang, Leroy, Meidt, McElroy, Pety, Rosolowsky, Saito, Sandstrom,
  Sancehz-Blazquez, \& Schruba}]{kreckel2019}
Kreckel, K., Ho, I.-T., Blanc, G.~A., {et~al.} 2019, ApJ, 887, 80,
  \dodoi{10.3847/1538-4357/ab5115}

\bibitem[{Kroupa(2001)}]{kroupa2001}
Kroupa, P. 2001, MNRAS, 322, 231, \dodoi{10.1046/j.1365-8711.2001.04022.x}

\bibitem[{Langeroodi {et~al.}(2023)Langeroodi, Hjorth, Chen, Kelly, Williams,
  Lin, Scarlata, Zitrin, Broadhurst, Diego, Huang, Filippenko, Foley, Jha,
  Koekemoer, Oguri, Perez-Fournon, Pierel, Poidevin, \&
  Strolger}]{langeroodi2023}
Langeroodi, D., Hjorth, J., Chen, W., {et~al.} 2023, ApJ, 957, 39,
  \dodoi{10.3847/1538-4357/acdbc1}

\bibitem[{Lara-L\'{o}pez {et~al.}(2010)Lara-L\'{o}pez, Cepa, Bongiovanni,
  {P\'{e}rez Garc\'{\i}a}, Ederoclite, {n}eda, {Fern\'{a}ndez Lorenzo},
  Povi\'{c}, \& S\'{a}nchez-Portal}]{laralopez2010}
Lara-L\'{o}pez, M.~A., Cepa, J., Bongiovanni, A., {et~al.} 2010, A\&A, 521,
  L53, \dodoi{10.1051/0004-6361/201014803}

\bibitem[{Leitherer {et~al.}(2014)Leitherer, Ekstr\"{o}m, Meynet, Schaerer,
  Agienko, \& Levesque}]{leitherer2014}
Leitherer, C., Ekstr\"{o}m, S., Meynet, G., {et~al.} 2014, ApJS, 212, 14,
  \dodoi{10.1088/0067-0049/212/1/14}

\bibitem[{Leitherer {et~al.}(2010)Leitherer, {Ortiz Ot\'{a}lvaro}, Bresolin,
  Kudritzki, {Lo Faro}, Pauldrach, Pettini, \& Rix}]{leitherer2010}
Leitherer, C., {Ortiz Ot\'{a}lvaro}, P.~A., Bresolin, F., {et~al.} 2010, ApJS,
  189, 309, \dodoi{10.1088/0067-0049/189/2/309}

\bibitem[{Lequeux {et~al.}(1979)Lequeux, Peimbert, Rayo, Serrano, \&
  Torres-Peimbert}]{lequeux1979}
Lequeux, J., Peimbert, M., Rayo, J.~F., Serrano, A., \& Torres-Peimbert, S.
  1979, A\&A, 80, 155

\bibitem[{Levesque {et~al.}(2010)Levesque, Kewley, \& Larson}]{levesque2010}
Levesque, E.~M., Kewley, L.~J., \& Larson, K.~L. 2010, AJ, 139, 712,
  \dodoi{10.1088/0004-6256/139/2/712}

\bibitem[{Li {et~al.}(2024)Li, Grasha, Krumholz, Wisnioski, Sutherland, Kewley,
  Chen, \& Li}]{li2024}
Li, S.-L., Grasha, K., Krumholz, M.~R., {et~al.} 2024, MNRAS, 529, 4993,
  \dodoi{10.1093/mnras/stae869}

\bibitem[{Lin(1989)}]{lin1989}
Lin, L. I.-K. 1989, Biometrics, 45, 255, \dodoi{10.2307/2532051}

\bibitem[{Luridiana {et~al.}(2015)Luridiana, Morisset, \& Shaw}]{luridiana2015}
Luridiana, V., Morisset, C., \& Shaw, R.~A. 2015, A\&A, 573, A42,
  \dodoi{10.1051/0004-6361/201323152}

\bibitem[{Maier {et~al.}(2006)Maier, Lilly, Carollo, Meisenheimer, Hippelein,
  \& Stockton}]{maier2006}
Maier, C., Lilly, S.~J., Carollo, C.~M., {et~al.} 2006, ApJ, 639, 858,
  \dodoi{10.1086/499518}

\bibitem[{Maiolino \& Mannucci(2019)}]{maiolino2019}
Maiolino, R., \& Mannucci, F. 2019, A\&ARv, 27, 3,
  \dodoi{10.1007/s00159-018-0112-2}

\bibitem[{Mannucci {et~al.}(2010)Mannucci, Cresci, Maiolino, Marconi, \&
  Gnerucci}]{mannucci2010}
Mannucci, F., Cresci, G., Maiolino, R., Marconi, A., \& Gnerucci, A. 2010,
  MNRAS, 408, 2115, \dodoi{10.1111/j.1365-2966.2010.17291.x}

\bibitem[{Mannucci {et~al.}(2021)Mannucci, Belfiore, Curti, Cresci, Maiolino,
  Marasco, Marconi, Mingozzi, Tozzi, \& Amiri}]{mannucci2021}
Mannucci, F., Belfiore, F., Curti, M., {et~al.} 2021, MNRAS, 508, 1582,
  \dodoi{10.1093/mnras/stab2648}

\bibitem[{Martin {et~al.}(2015)Martin, Drissen, \& Joncas}]{martin2015}
Martin, T., Drissen, L., \& Joncas, G. 2015, in Astronomical Society of the
  Pacific Conference Series, Vol. 495, Astronomical Data Analysis Software and
  Systems XXIV, ed. A.~R. Taylor \& E.~Rosolowsky, San Francisco, CA, 327

\bibitem[{Martin {et~al.}(2016)Martin, Prunet, \& Drissen}]{martin2016}
Martin, T.~B., Prunet, S., \& Drissen, L. 2016, MNRAS, 463, 4223,
  \dodoi{10.1093/mnras/stw2315}

\bibitem[{Mart\'{\i}n-Navarro {et~al.}(2015)Mart\'{\i}n-Navarro, Vazdekis, {La
  Barbera}, Falc\'{o}n-Barroso, Lyubenova, van~de Ven, Ferreras, S\'{a}nchez,
  Trager, Garc\'{\i}a-Benito, Mast, Mendoza, S\'{a}nchez-Bl\'{a}zquez,
  {Gonz\'{a}lez Delgado}, Walcher, \& {CALIFA Team}}]{martinnavarro2015}
Mart\'{\i}n-Navarro, I., Vazdekis, A., {La Barbera}, F., {et~al.} 2015, ApJL,
  806, L31, \dodoi{10.1088/2041-8205/806/2/l31}

\bibitem[{Masters {et~al.}(2014)Masters, McCarthy, Siana, Malkan, Mobasher,
  Atek, Henry, Martin, Rafelski, Hathi, Scarlata, Ross, Bunker, Blanc,
  Bedregal, Dom\'{\i}nguez, Colbert, Teplitz, \& Dressler}]{masters2014}
Masters, D., McCarthy, P., Siana, B., {et~al.} 2014, ApJ, 785, 153,
  \dodoi{10.1088/0004-637X/785/2/153}

\bibitem[{McCall {et~al.}(1985)McCall, Rybski, \& Shields}]{mccall1985}
McCall, M.~L., Rybski, P.~M., \& Shields, G.~A. 1985, ApJS, 57, 1,
  \dodoi{10.1086/190994}

\bibitem[{Meynet {et~al.}(1994)Meynet, Maeder, Schaller, Schaerer, \&
  Charbonnel}]{meynet1994}
Meynet, G., Maeder, A., Schaller, G., Schaerer, D., \& Charbonnel, C. 1994,
  A\&AS, 103, 97

\bibitem[{Mihalas(1972)}]{mihalas1972}
Mihalas, D. 1972, Non-LTE model atmospheres for B and O stars (Boulder, CO:
  National Center for Atmospheric Research)

\bibitem[{Mingozzi {et~al.}(2020)Mingozzi, Belfiore, Cresci, Bundy, Bershady,
  Bizyaev, Blanc, Boquien, Drory, Fu, Maiolino, Riffel, Schaefer,
  Storchi-Bergmann, Telles, Tremonti, Zakamska, \& Zhang}]{mingozzi2020}
Mingozzi, M., Belfiore, F., Cresci, G., {et~al.} 2020, A\&A, 636, A42,
  \dodoi{10.1051/0004-6361/201937203}

\bibitem[{Moll\'{a} {et~al.}(2009)Moll\'{a}, Garc\'{\i}a-Vargas, \&
  Bressan}]{molla2009}
Moll\'{a}, M., Garc\'{\i}a-Vargas, M.~L., \& Bressan, A. 2009, MNRAS, 398, 451,
  \dodoi{10.1111/j.1365-2966.2009.15160.x}

\bibitem[{Moll\'{a} {et~al.}(2006)Moll\'{a}, V\'{\i}lchez, Gavil\'{a}n, \&
  D\'{\i}az}]{molla2006}
Moll\'{a}, M., V\'{\i}lchez, J.~M., Gavil\'{a}n, M., \& D\'{\i}az, A.~I. 2006,
  MNRAS, 372, 1069, \dodoi{10.1111/j.1365-2966.2006.10892.x}

\bibitem[{Morisset {et~al.}(2016)Morisset, Delgado-Inglada, S\'{a}nchez,
  Galbany, Garc\'{\i}a-Benito, Husemann, Marino, Mast, \& Roth}]{morisset2016}
Morisset, C., Delgado-Inglada, G., S\'{a}nchez, S.~F., {et~al.} 2016, A\&A,
  594, A37, \dodoi{10.1051/0004-6361/201628559}

\bibitem[{Moustakas {et~al.}(2010)Moustakas, Kennicutt, Tremonti, Dale, Smith,
  \& Calzetti}]{moustakas2010}
Moustakas, J., Kennicutt, R.~C., Tremonti, C.~A., {et~al.} 2010, ApJS, 190,
  233, \dodoi{10.1088/0067-0049/190/2/233}

\bibitem[{Murphy {et~al.}(2011)Murphy, Condon, Schinnerer, Kennicutt, Calzetti,
  Armus, Helou, Turner, Aniano, {a}o, Bolatto, Brandl, Croxall, Dale, {Donovan
  Meyer}, Draine, Engelbracht, Hunt, Hao, Koda, Roussel, Skibba, \&
  Smith}]{murphy2011}
Murphy, E.~J., Condon, J.~J., Schinnerer, E., {et~al.} 2011, ApJ, 737, 67,
  \dodoi{10.1088/0004-637x/737/2/67}

\bibitem[{Nagao {et~al.}(2006)Nagao, Maiolino, \& Marconi}]{nagao2006}
Nagao, T., Maiolino, R., \& Marconi, A. 2006, A\&A, 459, 85,
  \dodoi{10.1051/0004-6361:20065216}

\bibitem[{Nicholls {et~al.}(2017)Nicholls, Sutherland, Dopita, Kewley, \&
  Groves}]{nicholls2017}
Nicholls, D.~C., Sutherland, R.~S., Dopita, M.~A., Kewley, L.~J., \& Groves,
  B.~A. 2017, MNRAS, 466, 4403, \dodoi{10.1093/mnras/stw3235}

\bibitem[{Osterbrock \& Ferland(2006)}]{osterbrock2006}
Osterbrock, D.~E., \& Ferland, G.~J. 2006, Astrophysics of gaseous nebulae and
  active galactic nuclei, 2nd edn. (Sausalito, CA: University Science Books)

\bibitem[{Pagel {et~al.}(1979)Pagel, Edmunds, Blackwell, Chun, \&
  Smith}]{pagel1979}
Pagel, B. E.~J., Edmunds, M.~G., Blackwell, D.~E., Chun, M.~S., \& Smith, G.
  1979, MNRAS, 189, 95, \dodoi{10.1093/mnras/189.1.95}

\bibitem[{Pauldrach {et~al.}(2001)Pauldrach, Hoffmann, \&
  Lennon}]{pauldrach2001}
Pauldrach, A. W.~A., Hoffmann, T.~L., \& Lennon, M. 2001, A\&A, 375, 161,
  \dodoi{10.1051/0004-6361:20021443}

\bibitem[{Peimbert {et~al.}(2017)Peimbert, Peimbert, \&
  Delgado-Inglada}]{peimbert2017}
Peimbert, M., Peimbert, A., \& Delgado-Inglada, G. 2017, PASP, 129, 082001,
  \dodoi{10.1088/1538-3873/aa72c3}

\bibitem[{Pellegrini {et~al.}(2020)Pellegrini, Rahner, Reissl, Glover, Klessen,
  Rousseau-Nepton, \& Herrera-Camus}]{pellegrini2020}
Pellegrini, E.~W., Rahner, D., Reissl, S., {et~al.} 2020, MNRAS, 496, 339,
  \dodoi{10.1093/mnras/staa1473}

\bibitem[{Peluso {et~al.}(2023)Peluso, Radovich, Moretti, Mingozzi, Vulcani,
  Poggianti, Marasco, \& Gullieuszik}]{peluso2023}
Peluso, G., Radovich, M., Moretti, A., {et~al.} 2023, ApJ, 958, 147,
  \dodoi{10.3847/1538-4357/acf833}

\bibitem[{P\'{e}rez-D\'{\i}az {et~al.}(2021)P\'{e}rez-D\'{\i}az, Masegosa,
  M\'{a}rquez, \& P\'{e}rez-Montero}]{perezdiaz2021}
P\'{e}rez-D\'{\i}az, B., Masegosa, J., M\'{a}rquez, I., \& P\'{e}rez-Montero,
  E. 2021, MNRAS, 505, 4289, \dodoi{10.1093/mnras/stab1522}

\bibitem[{P\'{e}rez-Montero(2014)}]{perezmontero2014}
P\'{e}rez-Montero, E. 2014, MNRAS, 441, 2663, \dodoi{10.1093/mnras/stu753}

\bibitem[{P\'{e}rez-Montero(2017)}]{perezmontero2017}
---. 2017, PASP, 129, 043001, \dodoi{10.1088/1538-3873/aa5abb}

\bibitem[{Petrosian {et~al.}(1972)Petrosian, Silk, \& Field}]{petrosian1972}
Petrosian, V., Silk, J., \& Field, G.~B. 1972, ApJ, 177, L69,
  \dodoi{10.1086/181054}

\bibitem[{Pilyugin(2005)}]{pilyugin2005_ff}
Pilyugin, L.~S. 2005, A\&A, 436, L1, \dodoi{10.1051/0004-6361:200500108}

\bibitem[{Pilyugin \& Grebel(2016)}]{pilyugin2016}
Pilyugin, L.~S., \& Grebel, E.~K. 2016, MNRAS, 457, 3678,
  \dodoi{10.1093/mnras/stw238}

\bibitem[{Pilyugin \& Thuan(2005)}]{pilyugin2005}
Pilyugin, L.~S., \& Thuan, T.~X. 2005, ApJ, 631, 231, \dodoi{10.1086/432408}

\bibitem[{Poetrodjojo {et~al.}(2019)Poetrodjojo, D'Agostino, Groves, Kewley,
  Ho, Rich, Madore, \& Seibert}]{poetrodjojo2019}
Poetrodjojo, H., D'Agostino, J.~J., Groves, B., {et~al.} 2019, MNRAS, 487, 79,
  \dodoi{10.1093/mnras/stz1241}

\bibitem[{Poetrodjojo {et~al.}(2018)Poetrodjojo, Groves, Kewley, Medling,
  Sweet, van~de Sande, Sanchez, Bland-Hawthorn, Brough, Bryant, Cortese, Croom,
  L\'{o}pez-S\'{a}nchez, Richards, Zafar, Lawrence, Lorente, Owers, \&
  Scott}]{poetrodjojo2018}
Poetrodjojo, H., Groves, B., Kewley, L.~J., {et~al.} 2018, MNRAS, 479, 5235,
  \dodoi{10.1093/mnras/sty1782}

\bibitem[{Polimera {et~al.}(2022)Polimera, Kannappan, Richardson, Bittner,
  Ferguson, Moffett, Eckert, Bellovary, \& Norris}]{polimera2022}
Polimera, M.~S., Kannappan, S.~J., Richardson, C.~T., {et~al.} 2022, ApJ, 931,
  44, \dodoi{10.3847/1538-4357/ac6595}

\bibitem[{Radovich {et~al.}(2019)Radovich, Poggianti, Jaff\'{e}, Moretti,
  Bettoni, Gullieuszik, Vulcani, \& Fritz}]{radovich2019}
Radovich, M., Poggianti, B., Jaff\'{e}, Y.~L., {et~al.} 2019, MNRAS, 486, 486,
  \dodoi{10.1093/mnras/stz809}

\bibitem[{Reddy {et~al.}(2023{\natexlab{a}})Reddy, Topping, Sanders, Shapley,
  \& Brammer}]{reddy2023a}
Reddy, N.~A., Topping, M.~W., Sanders, R.~L., Shapley, A.~E., \& Brammer, G.
  2023{\natexlab{a}}, ApJ, 952, 167, \dodoi{10.3847/1538-4357/acd754}

\bibitem[{Reddy {et~al.}(2023{\natexlab{b}})Reddy, Sanders, Shapley, Topping,
  Kriek, Coil, Mobasher, Siana, \& Rezaee}]{reddy2023b}
Reddy, N.~A., Sanders, R.~L., Shapley, A.~E., {et~al.} 2023{\natexlab{b}}, ApJ,
  951, 56, \dodoi{10.3847/1538-4357/acd0b1}

\bibitem[{Rosales-Ortega {et~al.}(2011)Rosales-Ortega, D\'{\i}az, Kennicutt, \&
  S\'{a}nchez}]{rosalesortega2011}
Rosales-Ortega, F.~F., D\'{\i}az, A.~I., Kennicutt, R.~C., \& S\'{a}nchez,
  S.~F. 2011, MNRAS, 415, 2439, \dodoi{10.1111/j.1365-2966.2011.18870.x}

\bibitem[{Rosales-Ortega {et~al.}(2012)Rosales-Ortega, S\'{a}nchez,
  Iglesias-P\'{a}ramo, D\'{\i}az, V\'{\i}lchez, Bland-Hawthorn, Husemann, \&
  Mast}]{rosalesortega2012}
Rosales-Ortega, F.~F., S\'{a}nchez, S.~F., Iglesias-P\'{a}ramo, J., {et~al.}
  2012, ApJL, 756, L31, \dodoi{10.1088/2041-8205/756/2/l31}

\bibitem[{Rousseau-Nepton {et~al.}(2018)Rousseau-Nepton, Robert, Martin,
  Drissen, \& Martin}]{rousseaunepton2018}
Rousseau-Nepton, L., Robert, C., Martin, R.~P., Drissen, L., \& Martin, T.
  2018, MNRAS, 477, 4152, \dodoi{10.1093/mnras/sty477}

\bibitem[{Rousseau-Nepton {et~al.}(2019)Rousseau-Nepton, Martin, Robert,
  Drissen, Amram, Prunet, Martin, Moumen, Adamo, Alarie, Barmby, Boselli,
  Bresolin, Bureau, Chemin, Fernandes, Combes, Crowder, {Della Bruna}, {Duarte
  Puertas}, Egusa, Epinat, Ksoll, Girard, {G\'{o}mez Llanos}, Gouliermis,
  Grasha, Higgs, Hlavacek-Larrondo, Ho, Iglesias-P\'{a}ramo, Joncas, Kam,
  Karera, Kennicutt, Klessen, lianou, Liu, Liu, de~Amorim, Lyman, Martel,
  Mazzilli-Ciraulo, McLeod, Melchior, Millan, Moll\'{a}, Momose, Morisset, Pan,
  Pati, Pellerin, Pellegrini, P\'{e}rez, Petric, Plana, Rahner, {Ruiz Lara},
  S\'{a}nchez-Menguiano, Spekkens, Stasi\'{n}ska, Takamiya, {Vale Asari}, \&
  V\'{\i}lchez}]{rousseaunepton2019}
Rousseau-Nepton, L., Martin, R.~P., Robert, C., {et~al.} 2019, MNRAS, 489,
  5530, \dodoi{10.1093/mnras/stz2455}

\bibitem[{S\'{a}nchez {et~al.}(2011)S\'{a}nchez, Rosales-Ortega, Kennicutt,
  Johnson, Diaz, Pasquali, \& Hao}]{sanchez2011}
S\'{a}nchez, S.~F., Rosales-Ortega, F.~F., Kennicutt, R.~C., {et~al.} 2011,
  MNRAS, 410, 313, \dodoi{10.1111/j.1365-2966.2010.17444.x}

\bibitem[{S\'{a}nchez {et~al.}(2015)S\'{a}nchez, P\'{e}rez, Rosales-Ortega,
  Miralles-Caballero, L\'{o}pez-S\'{a}nchez, iglesias P\'{a}ramo, Marino,
  S\'{a}nchez-Menguiano, Garc\'{\i}a-Benito, Mast, Mendoza, Papaderos, Ellis,
  Galbany, Kehrig, Monreal-Ibero, {Gonz\'{a}lez Delgado}, Moll\'{a}, Ziegler,
  de~Lorenzo-C\'{a}ceres, Mendez-Abreu, Bland-Hawthorn, Bekerait\.{e}, Roth,
  Pasquali, D\'{\i}az, Bomans, van~de Ven, \& Wisotzki}]{sanchez2015}
S\'{a}nchez, S.~F., P\'{e}rez, E., Rosales-Ortega, F.~F., {et~al.} 2015, A\&A,
  574, A47, \dodoi{10.1051/0004-6361/201424873}

\bibitem[{Sanders {et~al.}(2016)Sanders, Shapley, Kriek, Reddy, Freeman, Coil,
  Siana, Mobasher, Shivaei, Price, \& de~Groot}]{sanders2016}
Sanders, R.~L., Shapley, A.~E., Kriek, M., {et~al.} 2016, ApJ, 816, 23,
  \dodoi{10.3847/0004-637X/816/1/23}

\bibitem[{Sanders {et~al.}(2018)Sanders, Shapley, Kriek, Freeman, Reddy, Siana,
  Coil, Mobasher, Dav\'{e}, Shivaei, Azadi, Price, Leung, Fetherolf, de~Groot,
  Zick, Fornasini, \& Barro}]{sanders2018}
---. 2018, ApJ, 858, 99, \dodoi{10.3847/1538-4357/aabcbd}

\bibitem[{Sanders {et~al.}(2020)Sanders, Jones, Shapley, Reddy, Kriek, Coil,
  Siana, Mobasher, Shivaei, Price, Freeman, Azadi, Leung, Fetherolf, Zick,
  de~Groot, Barro, \& Fornasini}]{sanders2020}
Sanders, R.~L., Jones, T., Shapley, A.~E., {et~al.} 2020, ApJL, 888, L11,
  \dodoi{10.3847/2041-8213/ab5d40}

\bibitem[{Sanders {et~al.}(2021)Sanders, Shapley, Jones, Reddy, Kriek, Siana,
  Coil, Mobasher, Shivaei, Dav\'{e}, Azadi, Price, Leung, Freeman, Fetherolf,
  de~Groot, Zick, \& Barro}]{sanders2021}
Sanders, R.~L., Shapley, A.~E., Jones, T., {et~al.} 2021, ApJ, 914, 19,
  \dodoi{10.3847/1538-4357/abf4c1}

\bibitem[{Scheuermann {et~al.}(2023)Scheuermann, Kreckel, Barnes, Belfiore,
  Groves, Hannon, Lee, Minsley, Rosolowsky, Bigiel, Blanc, Boquien, Dale,
  Deger, Egorov, Emsellem, Glover, Grasha, Hassani, r.~Jeffreson, Klessen,
  Kruijssen, Larson, Leroy, Lopez, Pan, S\'{a}nchez-Bl\'{a}zquez, Santoro,
  Schinnerer, Thilker, Whitmore, Watkins, \& Williams}]{scheuermann2023}
Scheuermann, F., Kreckel, K., Barnes, A.~T., {et~al.} 2023, MNRAS, 522, 2369,
  \dodoi{10.1093/mnras/stad878}

\bibitem[{Shirazi {et~al.}(2014)Shirazi, Brinchmann, \& Rahmati}]{shirazi2014}
Shirazi, M., Brinchmann, J., \& Rahmati, A. 2014, ApJ, 787, 120,
  \dodoi{10.1088/0004-637x/787/2/120}

\bibitem[{Spearman(1904)}]{spearman1904}
Spearman, C. 1904, The Amer. Journ. of Psychology, 15, 72,
  \dodoi{10.1037/11491-005}

\bibitem[{Spitzer(1978)}]{spitzer1978}
Spitzer, L. 1978, Physical Processes in the Interstellar Medium (New York, NY:
  Wiley), \dodoi{10.1002/9783527617722}

\bibitem[{Stasi\'{n}ska(1990)}]{stasinska1990}
Stasi\'{n}ska, G. 1990, A\&AS, 83, 501

\bibitem[{Strom {et~al.}(2018)Strom, Steidel, Rudie, Trainor, \&
  Pettini}]{strom2018}
Strom, A.~L., Steidel, C.~C., Rudie, G.~C., Trainor, R.~F., \& Pettini, M.
  2018, ApJ, 868, 117, \dodoi{10.3847/1538-4357/aae1a5}

\bibitem[{Str\"{o}mgren(1939)}]{stromgren1939}
Str\"{o}mgren, B. 1939, ApJ, 89, 526, \dodoi{10.1086/144074}

\bibitem[{Sutherland {et~al.}(2018)Sutherland, Dopita, Binette, \&
  Groves}]{sutherland2018}
Sutherland, R., Dopita, M., Binette, L., \& Groves, B. 2018, MAPPINGS V:
  Astrophysical plasma modeling code, Astrophysics Source Code Library.
\newblock \url{http://ascl.net/1807.005}

\bibitem[{Thomas {et~al.}(2018)Thomas, Dopita, Kewley, Groves, Sutherland,
  Hopkins, \& Blanc}]{thomas2018}
Thomas, A.~D., Dopita, M.~A., Kewley, L.~J., {et~al.} 2018, ApJ, 856, 89,
  \dodoi{10.3847/1538-4357/aab3db}

\bibitem[{Thomas {et~al.}(2019)Thomas, Kewley, Dopita, Groves, Hopkins, \&
  Sutherland}]{thomas2019}
Thomas, A.~D., Kewley, L.~J., Dopita, M.~A., {et~al.} 2019, ApJ, 847, 100,
  \dodoi{10.3847/1538-4357/ab08a1}

\bibitem[{Tremonti {et~al.}(2004)Tremonti, Heckman, Kauffmann, Brinchmann,
  Charlot, White, Seibert, Peng, Schlegel, Uomoto, Fukugita, \&
  Brinkmann}]{tremonti2004}
Tremonti, C.~A., Heckman, T.~M., Kauffmann, G., {et~al.} 2004, ApJ, 613, 898,
  \dodoi{10.1086/423264}

\bibitem[{{Vale Asari} {et~al.}(2016){Vale Asari}, Stasi\'{n}ska, Morisset, \&
  {Cid Fernandes}}]{valeasari2016}
{Vale Asari}, N., Stasi\'{n}ska, G., Morisset, C., \& {Cid Fernandes}, R. 2016,
  MNRAS, 460, 1739, \dodoi{10.1093/mnras/stw971}

\bibitem[{van Zee {et~al.}(1998)van Zee, Salzer, Haynes, O'Donoghue, \&
  Balonek}]{vanzee1998}
van Zee, L., Salzer, J.~J., Haynes, M.~P., O'Donoghue, A.~A., \& Balonek, T.~J.
  1998, AJ, 116, 2805, \dodoi{10.1086/300647}

\bibitem[{Veilleux \& Osterbrock(1987)}]{veilleux1987}
Veilleux, S., \& Osterbrock, D.~E. 1987, ApJS, 63, 295, \dodoi{10.1086/191166}

\bibitem[{Vila-Costas \& Edmunds(1993)}]{vilacostas1993}
Vila-Costas, M.~B., \& Edmunds, M.~G. 1993, MNRAS, 265, 199,
  \dodoi{10.1093/mnras/265.1.199}

\bibitem[{V\'{\i}lchez {et~al.}(2019)V\'{\i}lchez, {n}o, Kennicutt, {De Looze},
  Moll\'{a}, \& Galametz}]{vilchez2019}
V\'{\i}lchez, J.~M., {n}o, M.~R., Kennicutt, R., {et~al.} 2019, MNRAS, 483,
  4968, \dodoi{10.1093/mnras/sty3455}

\bibitem[{Vilchez {et~al.}(1988)Vilchez, Pagel, Diaz, Terlevich, \&
  Edmunds}]{vilchez1988}
Vilchez, J.~M., Pagel, B. E.~J., Diaz, A.~I., Terlevich, E., \& Edmunds, M.~G.
  1988, MNRAS, 235, 633, \dodoi{10.1093/mnras/235.3.633}

\bibitem[{Virtanen {et~al.}(2020)Virtanen, Gommers, Oliphant, Haberland, Reddy,
  Cournapeau, Burovski, Peterson, Weckesser, Bright, van~der Walt, Brett,
  Wilson, Millman, Mayorov, Nelson, Jones, Kern, Larson, Carey, Polat, Feng,
  Moore, VanderPlas, Laxalde, Perktold, Cimrman, Henriksen, Quintero, Harris,
  Archibald, Ribeiro, Pedregosa, van Mulbregt, Vijaykumar, Bardelli, Rothberg,
  Hilboll, Kloeckner, Scopatz, Lee, Rokem, Woods, Fulton, Masson, Häggström,
  Fitzgerald, Nicholson, Hagen, Pasechnik, Olivetti, Martin, Wieser, Silva,
  Lenders, Wilhelm, Young, Price, Ingold, Allen, Lee, Audren, Probst, Dietrich,
  Silterra, Webber, Slavič, Nothman, Buchner, Kulick, Schönberger,
  de~Miranda~Cardoso, Reimer, Harrington, Rodríguez, Nunez-Iglesias,
  Kuczynski, Tritz, Thoma, Newville, Kümmerer, Bolingbroke, Tartre, Pak,
  Smith, Nowaczyk, Shebanov, Pavlyk, Brodtkorb, Lee, McGibbon, Feldbauer,
  Lewis, Tygier, Sievert, Vigna, Peterson, More, Pudlik, Oshima, Pingel,
  Robitaille, Spura, Jones, Cera, Leslie, Zito, Krauss, Upadhyay, Halchenko, \&
  Vázquez-Baeza}]{virtanen2020}
Virtanen, P., Gommers, R., Oliphant, T.~E., {et~al.} 2020, NatMe, 17, 261,
  \dodoi{10.1038/s41592-019-0686-2}

\bibitem[{Xiao {et~al.}(2018)Xiao, Stanway, \& Eldridge}]{xiao2018}
Xiao, L., Stanway, E.~R., \& Eldridge, J.~J. 2018, MNRAS, 477, 904,
  \dodoi{10.1093/mnras/sty646}

\bibitem[{Yeh \& Matzner(2012)}]{yeh2012}
Yeh, S. C.~C., \& Matzner, C.~D. 2012, ApJ, 757, 108,
  \dodoi{10.1088/0004-637x/757/2/108}

\bibitem[{Zhang {et~al.}(2017)Zhang, Yan, Bundy, Bershady, Haffner, Walterbos,
  Maiolino, Tremonti, Thomas, Drory, Jones, Belfiore, S\'{a}nchez,
  Diamond-Stanic, Bizyaev, Nitschelm, Andrews, Brinkmann, r.~Brownstein,
  Cheung, Li, Law, {Roman Lopes}, Oravetz, Pan, {Storchi Bergmann}, \&
  Simmons}]{zhang2017}
Zhang, K., Yan, R., Bundy, K., {et~al.} 2017, MNRAS, 466, 3217,
  \dodoi{10.1093/mnras/stw3308}

\bibitem[{Zou {et~al.}(2011)Zou, Zhang, Yang, Zhou, Jiang, Ma, Wu, Wu, Zhang,
  \& Fan}]{zou2011}
Zou, H., Zhang, W., Yang, Y., {et~al.} 2011, AJ, 142, 16,
  \dodoi{10.1088/0004-6256/142/1/16}

\bibitem[{Zovaro {et~al.}(2020)Zovaro, Sharp, Nesvadba, Kewley, Sutherland,
  Taylor, Groves, Wagner, Mukherjee, \& Bicknell}]{zovaro2020}
Zovaro, H. R.~M., Sharp, R., Nesvadba, N. P.~H., {et~al.} 2020, MNRAS, 499,
  4940, \dodoi{10.1093/mnras/staa3121}

\end{thebibliography}
\bibliographystyle{aasjournal.bst}

\end{document}